\documentclass[%
 reprint,
superscriptaddress,
showpacs,preprintnumbers,
nofootinbib,
 amsmath,amssymb,
 bm,
 aps,
]{revtex4-1}

\usepackage{graphicx}
\usepackage{amssymb}
\usepackage{epstopdf}
\usepackage{graphicx}
\usepackage{dcolumn}
\usepackage{bm}
\usepackage{color}
\usepackage[caption=false]{subfig}
\usepackage{mhchem}

\newcommand{\be}{\begin{eqnarray}}

\newcommand{\ee}{\end{eqnarray}}

\usepackage[normalem]{ulem}

\usepackage[export]{adjustbox}

\newcommand{\bl}{\boldsymbol{\ell}}
\newcommand{\dtl}[1]{\int \frac{\mathrm{d}^2\bl_{#1}'}{(2\pi)^2}}

\begin{document}

\preprint{APS/123-Q ED}

\title{Minimizing gravitational lensing contributions to the primordial bispectrum covariance}

\newcommand{\stockholm}{The Oskar Klein Centre for Cosmoparticle Physics,
Department of Physics, Stockholm University, SE-106 91 Stockholm, Sweden}
\newcommand{\princeton}{Department of Physics: Joseph Henry Laboratories, 
Jadwin Hall, Princeton University, 
Princeton, New Jersey 08542, USA}
\newcommand{\VSI}{Van Swinderen Institute for Particle Physics and Gravity,\\ University of Groningen,
Nijenborgh 4, 9747 AG Groningen, The Netherlands}
\newcommand{\kavli}{Institute of Astronomy and Kavli Institute for Cosmology, Madingley Road, Cambridge, UK, CB3 0HA}
\newcommand{\damtp}{DAMTP, Centre for Mathematical Sciences, Wilberforce Road, Cambridge, UK, CB3 0WA}
\newcommand{\imperial}{Physics Department, Imperial College London, Prince Consort Road, London SW7 2AZ, UK}
\newcommand{\cita}{Canadian Institute of Theoretical Astrophysics, 60 St George St, Toronto, ON M5S 3H8, Canada.}
\newcommand{\ASU}{School of Earth and Space Exploration, Arizona State University, Tempe, AZ, 85287, USA}

\author{William R. Coulton}
\affiliation{\kavli}
\author{P.\ Daniel Meerburg}
\affiliation{\VSI}
\author{David G. Baker}
\affiliation{\damtp}
\author{Selim Hotinli}
\affiliation{\imperial}
\author{Adriaan J.\ Duivenvoorden}
\affiliation{\princeton}
\author{Alexander van Engelen}
\affiliation{\ASU}
\affiliation{\cita}

\date{\today}

\begin{abstract}
The next generation of ground-based CMB experiments aim to measure temperature and polarization fluctuations up to $\ell_{\rm max} \approx 5000$ over half of the sky. Combined with {\it Planck} data on large scales, this will provide improved constraints on primordial non-Gaussianity. However, the impressive resolution of these experiments will come at a price. Besides signal confusion from galactic foregrounds, extra-galactic foregrounds and late-time gravitational effects, gravitational lensing will introduce large non-Gaussianity that can become the leading contribution to the bispectrum covariance through the connected 4-point function. Here, we compute this effect analytically for the first time on the full sky for both temperature and polarization. We compare our analytical results with those obtained directly from map-based simulations of the CMB sky for several levels of instrumental noise. Of the standard shapes considered in the literature, the local shape is most affected, resulting in a 35\% increase of the estimator standard deviation for an experiment like the Simons Observatory (SO) and a 110\% increase for a cosmic-variance limited experiment, including both temperature and polarization modes up to $\ell_{\rm max} = 3800$. Because of the nature of the lensing 4-point function, the impact on other shapes is reduced while still non negligible for the orthogonal shape. Two possible avenues to reduce the non-Gaussian contribution to the covariance are proposed. First by marginalizing over lensing contributions, such as the ISW-lensing 3pt function in temperature, and second by delensing the CMB. We show the latter method can remove almost all extra covariance, reducing the effect to below $<$5\% for local bispectra. At the same time, delensing would remove signal biases from secondaries induced by lensing, such as ISW-lensing. We aim to apply both techniques directly to the forthcoming SO data when searching for primordial non-Gaussianity. 
\end{abstract}

\pacs{Valid PACS appear here}
\maketitle

\section{Introduction}

The current cosmological paradigm suggests that the early Universe requires physics that generates Gaussian, isotropic and adiabatic initial fluctuations. As a result, the initial conditions of the Universe are completely characterised by the power spectrum (the harmonic equivalent of the 2-point correlation function) of scalar fluctuations $\zeta$, i.e.,
\be
\langle \zeta_{\vec{k}} \zeta_{\vec{k}'}\rangle &\equiv& (2 \pi)^3 \delta(\vec{k} + \vec{k}') P_{\zeta}(k).
\label{eq:2pt}
\ee
Further observational constraints have found that the power spectrum is well described by
\be
P_{\zeta}(k) &=& 2\pi^2 A_s k^{-3}\left(\frac{k}{k_*}\right)^{n_s-1},
\label{eq:PS}
\ee
where $A_s$ is the amplitude of initial fluctuations and $n_s$ is the scale dependence of these fluctuations. There are numerous theoretical models for their origin that are compatible with the current, and any near future, stringent observational constraints on the power law above (see e.g. \citep{Planck2018X}). It has long been realized that additional observational constraints are required to make empirical progress in the understanding of the early Universe. One such avenue is through measurements of primordial non-Gaussianity. Primordial non-Gaussianity introduces moments beyond the 2-point correlation function of Eq.~\eqref{eq:2pt} and has the potential to provide critical new information on many aspects relevant to the physics in the early Universe. Theoretically the most interesting and observationally the most studied observable is the primordial 3-point function, or its harmonic-space equivalent - the bispectrum \citep{Spergel1999}. Numerous studies have been performed, both on the theoretical side (see e.g. \cite{Meerburg2019} and references therein), exploring different models of the early Universe, and the observational side, developing efficient and optimal estimators \citep[e.g.][]{Komatsu2002,Komatsu2005,Yadav2007,Fergusson2010,Smith2011,Bucher2016,Munchmeyer:2014cca}. Although no evidence exists that the early Universe requires modelling beyond the simple power spectrum of Eq.~\eqref{eq:PS}, the bispectrum, and in particular its amplitude $f_{\rm NL}$, remains one of the most illustrious targets in cosmology. Compared to the power spectrum, the bispectrum has many more degrees of freedom besides its amplitude. Generally, the 3-point function of primordial scalar fluctuations is written as \cite{Acquaviva:2002ud,Maldacena2003,Creminelli:2003iq, Meerburg:2009ys} 
\be
\langle \zeta_{\vec{k}} \zeta_{\vec{k}'} \zeta_{\vec{k}''}\rangle &\equiv& (2 \pi)^3 \delta(\vec{k} + \vec{k}' + \vec{k}'') B(k,k',k''),
\ee
where 
\be
B(k,k',k'') = f_{\rm NL} A(A_s) S(k,k',k'') /(kk'k'')^2 .
\ee
Here $A(A_s)$ is a normalization function depending only on the amplitude of the scalar power spectrum $A_s$ \cite{Planck2019IX} and $S$ is typically referred to as the shape of the bispectrum \cite{Babich2004a}. In a statistically isotropic universe, the power spectrum is only a function of the amplitude of the comoving wavevector $k$ which, with scale invariance, ensures that (to good approximation) $P_{\zeta}(k) \propto k^{-3}$. The bispectrum, on the other hand, even in an isotropic universe, is a free function of 3 momenta that form a connected triangle in Fourier space. 
Theoretical motivations, as well as observational limitations, have resulted in reducing the functional complexity of the bispectrum, $B$, into well-defined `shapes'. These different shapes can discriminate between different theoretical models while the restriction to considering a few shapes also reduces the computational challenge (as measuring the full bispectrum is intractable). The most well known and best studied shapes are the local, equilateral and orthogonal shape \citep{Gangui2000,Komatsu2001,Babich2004a,Chen2007,Meerburg2009,Senatore2010}. Each of these shapes has been proposed to be able to identify a specific characteristic of the early Universe, contextualised by inflation (see e.g. Ref.~\cite{Chen2010} for a review and references in \citep{Meerburg2019} for recent developments). {\it Planck} has put the most stringent constraints on these bispectra (among many others)~\citep{Planck2019IX}. Thus far, the data suggests that there is no evidence for a non-zero bispectrum. However, theoretically, it is possible to identify interesting thresholds for these shapes that separate different classes of early universe models such as $f_{\rm NL} \sim 1$, which separates between single- and many-field inflationary models \citep{Creminelli2003,Maldacena2003}. These thresholds have not yet been reached and it has been shown that it will generally be quite a challenge to reach this threshold in the near future from measurements of the bispectrum. As it stands, the local shape is the most likely to be constrained to this level in the next decade\footnote{Although possibly not by measuring the bispectrum alone. See e.g. \cite{Dalal2008,McDonald2009,Munchmeyer2019}} \citep{Dore2014}, while other shapes will require more futuristic observations to reach this threshold \citep{Karagiannis2018,Meerburg2019}. Whilst such thresholds provide clear targets for future missions, theoretically the window between current bounds and these thresholds is still wide-open, which motivates a continued search. 

Observationally, the CMB is still the cleanest observable to constrain non-Gaussianities, as, to good approximation, the temperature and polarization fluctuations in the CMB can be described by linear physics. Hypothetically, any observed level of non-Gaussianity in the (primary) CMB should therefore be sourced primordially. Historically, the linear approximation of the CMB has been sufficient to constrain most shapes of primordial non-Gaussianity. However, as the resolution of CMB experiments improves, this assumption will no longer be valid and addressing non-linearities will be required in order to extract the underlying primordial signal. The effects that imprint themselves on the CMB after last scattering are known as CMB secondary anisotropies and will produce their own non-Gaussianity. In this work, the most important secondary anisotropies are those arising from the gravitational lensing of CMB photons by the intervening large scale structure, which deflects photon paths by a few arcminutes, but in a way that is coherent across several degrees. The result is that CMB modes that were initially statistically independent become correlated through lensing. This induced correlation is useful as it allows us to reconstruct maps of the lensing field and extract cosmological information \citep{Okamoto2003,Das2011,PlanckVIII2018,Wu2019}. However, this lensing-induced correlation also reduces the number of independent CMB modes to average over when performing estimates of the properties of the primordial CMB sky, such as non-Gaussianity. 

In the context of bispectrum analyses, a well-studied example of the importance of CMB lensing is through the ISW-lensing effect \citep{Goldberg1999}, which generates a non-zero bispectrum due to the correlation between lensing of the CMB on small scales and the late-time integrated Sach-Wolfe effect on large scales. Both effects are sourced by the gravitational potential, and their correlation produces a well-known non-zero signal. Because of the gravitational nature of the effect, it has a local shape, i.e., long modes are correlated with small scale power. For that reason, the ISW-lensing bispectrum is primarily a nuisance for a primordial signal of the local type \citep{Hanson2009,Smith2011,Mangilli2009}. The polarised component of the CMB does not have an ISW signal, but instead has a reionization signal on large angular scales \citep{Cooray2000}. Although this does introduce a non-zero bispectrum, its amplitude is much smaller \citep{Lewis2011}. Besides the ISW-lensing effect, many other bispectra are sourced by CMB secondaries. A recent study \citep{Hill2018} has shown that some of these bispectra might exceed the lensing induced bispectra. Due to the uncertainties in the astrophysical processes that source these bispectra, their impact and amplitude is somewhat uncertain and these are the topic of ongoing investigations. Besides secondaries, galactic foregrounds will also introduce non-zero bispectra \citep{Komatsu2002,Jung2018,Coulton2019}. Some of these can be dealt with by proper frequency cleaning of the CMB maps, but we should keep in mind the potential smallness of the primordial signal, which can easily be obscured by tiny residual foregrounds. 

So far the discussion has revolved around signal confusion, also known as biases to the measured non-Gaussianity. However the covariance of these measurements will also be impacted by non-linear effects. Previously, lensing has been shown to affect the covariance of the two-point power spectra as well as the covariance between the two-point power spectra and the four-point-based lensing estimates \citep{Smith2004,Smith2006,Li2007,BenoitLevy2012,Schmittfull2013,Peloton2017}. This work extends this analysis to the bispectrum.

It is not uncommon in the CMB community to assume, when constraining primordial non-Gaussianity, that the largest contribution to the variance of the bispectrum comes from the disconnected part of the 6-point function, i.e., the variance is dominated by products of 3 power spectra. However, it was realized in Ref.~\cite{Babich2004b} that for resolutions beyond {\it Planck} the connected parts of the 6-point function induced by lensing will quickly dominate the contributions from the disconnected part. Since Ref.~\cite{Babich2004b}, no study has been performed (to our knowledge) that addresses the effects of this on primordial non-Gaussianity searches, thoughthis effect was included in Ref.~\cite{Lewis2011} when they examined the detectability of the ISW-lensing bispectrum. As the sensitivity of high resolution CMB experiments is further pushed, and we hope to use this data to further constrain or detect primordial non-Gaussianity, it is timely to establish the impact of this effect in more detail with future data in mind. 

In this paper we recompute the lensing induced covariance of the observational bispectrum. For the first time, we do this on the full sky. Because of the nature of the effect (lensing) the largest contributions are expected on large scales (correlated with small scale power). By working on the full sky we ensure that we are accurately capturing the induced extra covariance. We compare our result to those presented in Ref.~\cite{Babich2004b}. Theoretically, we limit ourselves to contributions linear in the lensing power spectrum. Interestingly, it was shown in e.g.~\cite{Schmittfull2013} that for induced lensing covariance on measurements of the power spectrum, terms quadratic in the lensing power spectrum are more important. We compare our results to simulations by applying binned and KSW bispectrum estimators to lensed maps \citep{Komatsu2005,Bucher2010} and find that our linear approximation is sufficient. In the remainder of the paper we explore ways to deal with this extra covariance. We investigate two avenues. First, we marginalize over known lensing contributions to the CMB bispectrum, such as the ISW-lensing bispectrum. Theoretically, this `new' estimator should contain less lensing contributions and its lensing induced covariance should be reduced. Second, before applying bispectrum estimators to the data, we try to remove the lensing effect by delensing. Delensing was initially introduced to remove the lensing induced $B$ modes in order to aid the search for evidence of primordial gravitational waves \citep{Smith2012,Marian2007}. For temperature and $E$-mode polarization, delensing has been shown to be useful in reducing the lensing contributions to the 2-point covariance \citep{Green2017}. Here, delensing is applied to remove excess bispectrum covariance. We end by discussing the practical advantages and limitations of both techniques and comment on applying proposed methods to upcoming CMB data. 

\section{The CMB Bispectrum}

The CMB bispectrum describes the $3$-point correlation of temperature and polarization anisotropies in the harmonic domain. The harmonic representation means that the temperature and polarization anisotropies are described in terms of their spherical harmonic coefficients. Before presenting the bispectrum, we therefore briefly introduce these harmonic coefficients.

The temperature harmonic modes are defined as follows:
\begin{align}
a^T_{\ell m} = \int_{S^2} {\mathrm{d}^2\bm{\hat{n}}} \, T(\bm{\hat{n}}) \, Y^*_{\ell m}(\bm{\hat{n}}) \, ,
\end{align}
where $T(\bm{\hat{n}})$ denotes the 
CMB temperature at position $\bm{\hat{n}} \in S^2$ on the celestial sphere. In terms of the standard spherical coordinates, the differential $\mathrm{d}^2 \bm{\hat{n}}$ is given by $\sin \theta \ \mathrm{d}\phi \ \mathrm{d} \theta$ and $Y_{\ell m}$ is a spherical harmonic of degree $\ell$ and order $m$. On sufficiently small scales we can rewrite the above using the flat-sky approximation: 
\be
a^T_{\ell m} \rightarrow T(\bl) \equiv \int {\mathrm{d}^2\bm{\hat{n}}}{} \, T(\bm{\hat{n}}) e^{-i \bl \cdot \bm{\hat{n}}} \, ,
\ee 
where $\bl$ is now a continuous 2-dimensional vector and $\bm{\hat{n}}$ is in the plane of the sky. 

The linearly polarized component of the CMB is described by two (real) fields: $Q(\bm{\hat{n}})$ and $U(\bm{\hat{n}})$. These fields represent the elements of a symmetric, traceless tensor field and are therefore coordinate-dependent quantities that transform among themselves under coordinate rotations. For that reason, it is convenient to combine these fields into two complex `spin-$2$' fields on the sphere that are defined as follows: 
 \begin{align}
{}^{(\pm 2)}P(\bm{\hat{n}}) \equiv (Q\pm iU)(\bm{\hat{n}}) \, ,
\end{align}
and transform as: 
\begin{align}\label{eq:p_spin_transform}
{}^{(\pm 2)}P(\bm{\hat{n}}) \mapsto {}^{(\pm 2)}P(\bm{\hat{n}}) e^{\mp 2 i \psi} \, ,
\end{align}
under a right-handed rotation by an angle $\psi$ of the local coordinate system around the direction $\bm{\hat{n}}$. 

The harmonic decomposition of ${}^{(\pm 2)}P$ is performed using the spin-weighted spherical harmonics ${}_sY_{\ell m}$ with spin weights $s=\pm2$~\citep{Zaldarriaga:1996xe}. The resulting harmonic coefficients conveniently lose the implicit coordinate dependence that ${}^{(\pm 2)}P$ has, but still mix under parity. For that reason it is convenient instead to use the $E$- and $B$-modes: two linear combinations of the harmonic coefficients that behave as parity eigenstates ~\citep{Zaldarriaga:1996xe, Kamionkowski:1996zd}. The $E$- and $B$-modes are related to the ${}^{(\pm 2)}P$ field as follows:
\begin{align} \label{eq_alm_EB}
\begin{split}
a^E_{\ell m} &= - \frac{1}{2}\sum\limits_{s \in \pm2} \int_{S^2} \! \! \mathrm{d}^2\bm{\hat{n}} \, {}^{(s)}P(\bm{\hat{n}}) \, {}_{s}Y^*_{\ell m}(\bm{\hat{n}}) \, , \\
a^B_{\ell m} &= - \frac{1}{2i} \sum\limits_{s\in \pm2} \! \! \mathrm{sgn}(s) \! \! \int_{S^2} \! \! \mathrm{d}^2\bm{\hat{n}}\, {}^{(s)}P(\bm{\hat{n}}) \, {}_{s}Y^*_{\ell m}(\bm{\hat{n}}) \, .
\end{split}
\end{align}
Under parity, the odd moments of the $E$-mode field gain a minus sign. The $B$-mode field shows the opposite behaviour:
\begin{align}\label{eq:parity}
\begin{split} 
a^E_{ \ell m} &\mapsto (-1)^{\ell} a^E_{ \ell m} \,, \\
a^B_{ \ell m} &\mapsto (-1)^{\ell+1} a^B_{ \ell m} \, .
\end{split}
\end{align}
Note that the harmonic coefficients of the temperature anisotropies transform as the $E$-modes above. 

When the flat-sky approximation is applied to the $Q$ and $U$ fields, the definitions of the $E$- and $B$-modes have to be replaced by the following \citep{Hu2000}:
\begin{align}
a^E_{\ell m} &\rightarrow E(\bl) \equiv Q(\bl) \cos(2\psi_\ell) +U(\bl)\sin(2\psi_\ell)\, \\
a^B_{\ell m} &\rightarrow B(\bl) \equiv -Q(\bl) \sin(2\psi_\ell) +U(\bl)\cos(2\psi_\ell) \, ,
\end{align} 
where here $\bl$ is a vector in the 2D Fourier plane, and $\psi_\ell$ is the angle formed between $\bl$ and the positive $\ell_x$ axis.

With the harmonic modes defined, we now move on to the bispectrum. As mentioned before, the bispectrum is the harmonic equivalent of the three point function and is defined as
\begin{align}
B_{m_1 m_2 m_3}^{ X_1 X_2 X_3,\ell_1 \ell_2 \ell_3} \equiv \big\langle a^{X_1}_{\ell_1 m_1} a^{X_2}_{\ell_2 m_2} a^{X_3}_{\ell_3 m_3}\big\rangle \, ,
\end{align}
with $X_1, X_2, X_3 \in \{T, E, B\}$. 
Under the assumption of statistical isotropy, the bispectrum can be factored into a Wigner 3-$j$ symbol and the angle-averaged bispectrum~\citep{Spergel1999, Hu2001}: 
\begin{align}\label{eq:isotropy_b}
B_{m_1 m_2 m_3}^{X_1 X_2 X_3,\ell_1 \ell_2 \ell_3} = 
\begin{pmatrix}\ell_1 & \ell_2 & \ell_3 \\ m_1 & m_2 & m_3 \end{pmatrix} B_{\ell_1 \ell_2 \ell_3}^{X_1 X_2 X_3} \, .
\end{align}
Note that the orthogonality of the $3$-$j$ symbols~\citep{edmonds_1957} implies that that above relation may be inverted; the angle-averaged bispectrum of a (noiseless) dataset $a^X_{\ell m}$ can therefore be estimated as follows:
\be
\widehat{B}_{\ell_1,\ell_2,\ell_3}^{X_1,X_2,X_3} =\sum\limits_{m_1,m_2,m_3} \left( \begin{matrix} \ell_1 & \ell_2 & \ell_3 \\ m_1 & m_2 & m_3 \end{matrix} \right) a^{X_1}_{\ell_1 m_1} a^{X_2}_{\ell_2 m_2} a^{X_3}_{\ell_3 m_3}.\nonumber \\
\ee

The flat-sky equivalent of the bispectrum is defined as follows:
\be
 &B^{X_1,X_2,X_3}(\bl_1,\bl_2,\bl_3) \equiv \langle X(\bl_1) X(\bl_2) X(\bl_3)\rangle 
\ee
In the flat sky the analog to the angle averaged bispectrum is the reduced bispectrum, $b(\ell_1,\ell_2,\ell_3)$ which is defined as
\begin{align}
&B^{X_1,X_2,X_3}(\bl_1,\bl_2,\bl_3) \nonumber \\
&= 4\pi^2 \delta^{(2)}(\bl_1+\bl_2+\bl_3)b^{X_1,X_2,X_3}(\ell_1,\ell_2,\ell_3),
\end{align}
where we have assumed isotropy and parity invariance in the second line. The flat-sky reduced bispectrum is related to the full-sky angle-averaged bispectrum as~\citep{Hu2000}:
\be
B_{\ell_1 \ell_2 \ell_3}^{X_1 X_2 X_3}= I(\ell_1,\ell_2,\ell_3) b^{X_1,X_2,X_3}(\ell_1, \ell_2, \ell_3)
\ee
where $I(\ell_1,\ell_2,\ell_3)$ is given by
\be
I(\ell_1,\ell_2,\ell_3) &=&\sqrt{\frac{(2\ell_1+1)(2\ell_2+1)(2\ell_3+1)}{4\pi}}\left( \begin{matrix} \ell_1 & \ell_2 & \ell_3 \\ 0 & 0 & 0 \end{matrix} \right) \nonumber. 
\ee
To maintain generality for parity even and parity odd cases (and to express the equations cleanly), we begin with a discussion of the variance of the full bispectrum estimator in the flat sky
\begin{align}
 \widehat{B}^{X_1,X_2,X_3}&(\bl_1,\bl_2,\bl_3) =  \nonumber \\ &4\pi^2 \delta^{(2)}(\bl_1+\bl_2+\bl_3) X(\bl_1) X(\bl_2) X(\bl_3) .
\end{align}
We note that in practice this estimator would not be used and instead an estimator of the form described in Section \ref{sec:fnlEstimators} would be used. However our discussion of the full bispectrum applies equally to these estimators, as is discussed in Section \ref{sec:fnlEstimators}.

\section{Gaussian covariance}
Let us start by re-deriving the covariance in the weak non-Gaussianity limit, where contributions beyond the power spectrum can be neglected. We consider $X = T$ and drop this index for clarity. We have 
\be
\langle \widehat{B}_{\ell_1 \ell_2 \ell_3} \widehat{B}^*_{\ell_1' \ell_2' \ell_3'}\rangle &=&\sum\limits_{mm'} \left( \begin{matrix} \ell_1 & \ell_2 & \ell_3 \\ m_1 & m_2 & m_3 \end{matrix} \right) \left( \begin{matrix} \ell_1' & \ell_2' & \ell_3' \\ m_1' & m_2' & m_3'\end{matrix} \right) \nonumber \\
&& \langle a_{\ell_1 m_1} a_{\ell_2 m_2} a_{\ell_3 m_3} a^*_{\ell_1' m_1'} a^*_{\ell_2' m_2'} a^*_{\ell_3' m_3'}\rangle. \nonumber \\ 
\ee
To proceed we use the identity 
\be
\sum\limits_{m_1 m_2}\left( \begin{matrix} \ell_1 & \ell_2 & L \\ m_1 & m_2 & M \end{matrix} \right) \left( \begin{matrix} \ell_1 & \ell_2 & L' \\ m_1 & m_2 & M'\end{matrix} \right) &=& \frac{\delta_{LL'}\delta_{MM'}}{2L+1}
\ee
and the definition of the power spectrum 
\be
\langle a_{\ell m} a^*_{\ell'm'} \rangle = C_{\ell} \delta_{\ell \ell'} \delta_{m m'}.
\ee
Note that in our notation $C_{\ell}$ is the observed power spectrum and thus includes the effects of the beam, noise and galactic and extragalactic foregrounds, and $\tilde{C}_{\ell}$ is the unlensed CMB spectrum. 
Using these we find 
\be
\langle \widehat{B}_{\ell_1 \ell_2 \ell_3} \widehat{B}_{\ell_1' \ell_2' \ell_3'}\rangle &=& C_{\ell_1} C_{\ell_2} C_{\ell_3} \left[\delta_{\ell_1 \ell_1'} \delta_{\ell_2 \ell_2'} \delta_{\ell_3 \ell_3'} \right.\nonumber \\
&& \left.+\delta_{\ell_1 \ell_2'} \delta_{\ell_2 \ell_3'} \delta_{\ell_3 \ell_1'}+\delta_{\ell_1 \ell_3'} \delta_{\ell_2 \ell_1'} \delta_{\ell_3 \ell_2'} \right. \nonumber \\
&& \left.+ \delta_{\ell_1 \ell_2'} \delta_{\ell_2 \ell_1'} \delta_{\ell_3 \ell_3'}+\delta_{\ell_1 \ell_3'} \delta_{\ell_2 \ell_2'} \delta_{\ell_3 \ell_1'}+ \right. \nonumber \\
&& \left. \delta_{\ell_1 \ell_1'} \delta_{\ell_2 \ell_3'} \delta_{\ell_3 \ell_2'} \right]
\ee
The diagonal then becomes 
\be
\langle \widehat{B}_{\ell_1 \ell_2 \ell_3}^2\rangle &=& \prod_{i=1}^{3} C_{\ell_i} \left[1+2\delta_{\ell_1 \ell_2} \delta_{\ell_2 \ell_3}+\delta_{\ell_1\ell_2} + \delta_{\ell_2 \ell_3} + \delta_{\ell_2 \ell_1}\right] \nonumber, 
\label{eq:GCov}
\ee
i.e., a familiar result where the variance is enhanced by a factor of 2 if two $\ell$'s are equal and a factor of 6 when all $\ell$'s are equal. 

The variance of the bispectrum in the flat sky is given by an identical equation with the dirac deltas replaced with delta functions.

\section{Effect of lensing}\label{sec:effectOfLensing}

In the derivation above we explicitly assumed that, when expanding the 6-point function, the only remaining contributions come from the connected 2-point functions. If the observed CMB was completely described by a linear transformation of the initial fluctuation $\zeta$, and contained only primordial non-Gaussianity, this assumption would be a good approximation (i.e., the weak non-Gaussian limit). However, on small scales the observed CMB is no longer linear. Lensing is a second-order effect that introduces a correction to the statistical properties the CMB and produces non-zero connected $n$-point functions for $n>2$. First, in the derived weak non-Gaussianity limit above, the temperature spectra $C_{\ell}$ should be replaced with their lensed versions. Second, and this turns out to be more important, lensing introduces large non-Gaussianity, which appears first in the connected 4-point function. In other words, lensing will modify the covariance as~\citep{Coulton2018,Kayo2013,Duivenvoorden2019}:
\be \label{eq:varComps}
&{\rm Var}(\widehat{B}) = {\rm Var}(\widehat{B})_{G} + &\Delta {\rm Var}(\widehat{B})_{G} +\Delta {\rm Var}(\widehat{B})_{\rm connected}, \nonumber \\
\ee
Here ${\rm Var}(\widehat{B})_{G}$ is the covariance in the absence of lensing, $\Delta {\rm Var}(\widehat{B})_{G} $ captures changes in the variance arising due to lensing altering the power spectrum and $\Delta {\rm Var}(\widehat{B})_{\rm connected}$ contains the non-Gaussian contributions from lensing to the covariance. The last term includes all the contributions which cannot be expressed as a product of three power spectra. The connected contributions can be broken down into 
\be
&\Delta {\rm Var}(\widehat{B})_{\rm connected}= & {\rm Var}(\widehat{B})_{2\times3p}+{\rm Var}(\widehat{B})_{2p\times4p}  \nonumber \\&&+ {\rm Var}(\widehat{B})_{6p}.
\label{eq:LensingCov}
\ee
The first term on the right arises from a product of two bispectra and is caused by terms like the aforementioned ISW-lensing bispectrum. However, for our purposes these corrections are subdominant to the other corrections. The second and third terms introduce the connected 4-and-6-point functions from lensing. To lowest order in the lensing potential, $\phi$, the connected 4-point function will be $\propto C_{\ell''} C_{\ell} C^{\phi\phi}_{\ell'}$. In other words, the connected contribution term will start dominating the covariance when $K(\ell,\ell') C_{\ell} C_{\ell'}^{\phi\phi} \geq C_{\ell} $. Here $K(\ell,\ell')$ presents a coupling kernel\footnote{As we will later explain, we drop `loop' corrections in this simple picture, which could introduce internal summations betweem the kernel and the power spectra.} that determines the strength of the coupling between the temperature/polarization and the lensing power spectra. Note that this coupling kernel plays an important role. If lensing was purely a perturbative effect, then for it to become the dominant contribution to the covariance would require $C^{\phi \phi}_{\ell} \geq 1$, i.e., fluctuations in the gravitational potential need to be of $\mathcal{O}(1)$. Instead, the coupling kernel assures that perturbations are still under control, but the effect can become large as many configurations are summed over. This can occur if the kernels couples scales in a manner similar to the signal from primordial non-Gaussianity. In fact, the kernel strongly couples large scale modes and small scale power, which is very similar to the coupling produced by primordial non-Gaussianity of the local type. Thus we find that local non-Gaussianity searches are most affected by lensing, while other shapes are less affected (they are obviously affected by the first term on the RHS of Eq.~\eqref{eq:LensingCov}). Finally, we expect that lensing contributions from the 6-point function, which arise at higher order in $C^{\phi\phi}$, will be subdominant.

\subsection{Flat-sky Temperature}
First we compute the lensing induced covariance in the flat sky. The obtained results are only valid for $\ell \gtrsim 10$ and accurate results are expected only when $\ell \gtrsim 40$ (as we do not use the extended Limber approximation here \citep{Limber1953,Loverde2008}). The purpose of calculating the flat-sky bispectrum covariance is two-fold. First, it allows us to compare our results to those obtained in Ref \cite{Babich2004b}. Second, even the simplest term in the final expression for the bispectrum covariance depends on 5 degrees of freedom. In the flat sky, it is possible to use both angles (2) and multipoles (3), while in the full sky these are strictly multipoles (5). For high resolution, e.g. $\ell_{\rm max} \sim 4000$, the effect on the full-sky covariance, which requires summing over these degrees of freedom, becomes a computational challenge whilst the flat-sky computation is more feasible. It is useful to be able to compare the full-sky calculations to the flat sky in a range of multipoles where they are both valid and then use the flat-sky results to extend out to higher multipoles. In addition, more complicated terms involve an additional sum in the full sky and contain a Wigner 6-$j$ symbol. As a result, these terms are even harder to compute numerically. Fortunately, these `loop corrections' are small, but their smallness is easier to check in the flat sky (as they are computationally tractable). 

In the flat sky, the power spectra are given by 
\be
\langle \tilde{T}(\bl) \tilde{T}^*(\bl')\rangle &=& (2\pi)^2 \delta(\bl-\bl') \tilde{C}_{\ell},\\
\langle T(\bl) T^*(\bl')\rangle &=& (2\pi)^2 \delta(\bl-\bl') C_{\ell},\\
\langle \phi(\bl) \phi^*(\bl')\rangle &=& (2\pi)^2 \delta(\bl-\bl') C^{\phi\phi}_{\ell},
\ee
where $ \tilde{T}(\bl)$ is the unlensed CMB temperature field.
In the flat sky the lensed temperature field can be written as \citep{Hu2000}
\be
T(\bl) = \tilde{T}(\bl) - \dtl{1}\tilde{T}(\bl'_1)\phi(\bl-\bl'_1)(\bl-\bl'_1)\cdot\bl'_1,\nonumber\\
\ee
to linear order in $\phi$. Note that whilst the correction to the variance from power spectrum changes (the second term in Eq. \eqref{eq:varComps}) requires expanding the fields to second order in $\phi$, the connected term, which is the focus of this paper, can be computed to the same order in $C^{\phi \phi}$ by expanding only to first order in $\phi$.

First, let us define the total covariance in the flat sky
\begin{widetext}
\be
{\rm Var}(\widehat{B}) \equiv \langle \widehat{B} \widehat{B}^* \rangle = (2\pi)^2\delta^{(2)}(\sum\limits_i \bl_i) (2\pi)^2\delta^{(2)}(\sum\limits_i\bl_i') \langle T(\bl_1)T(\bl_2)T(\bl_3)T^*(\bl_1')T^*(\bl_2')T^*(\bl_3')\rangle.
\ee

Next, we compute the ${\rm Var}(\widehat{B})_{2p\times4p}$ contribution to the bispectrum covariance. The details of this calculation can be found in Appendix \ref{app:flatSkyderivation}. We find
\be \label{eq:flatSky2x4pnt}
{\rm Var}(\widehat{B})_{2p\times4p} &=& (2\pi)^8\delta^{(2)}(0)\delta^{(2)}(\bl_1-\bl_1')\delta^{(2)}(\sum\limits_i \bl_i)\delta^{(2)}(\sum\limits_i \bl_i') C_{\ell_1} C_{\ell_1}^{\phi\phi} \bl_1\cdot \bl_2 \tilde{C}_{\ell_2} \bl_1\cdot \bl_2' \tilde{C}_{\ell_2'} +{\rm 35\;perm.} + \nonumber \\
& & (2\pi)^6\delta^{(2)}(\bl_1-\bl_1') \delta^{(2)}(\sum\limits_i \bl_i)\delta^{(2)}(\sum\limits_i \bl_i')C_{\ell_1} \int \mathrm{d}^2\bm{\hat{n}} C_{|\bl_2-\bl_3'|}^{\phi\phi} G(\bl_2,\bl_3')\tilde{C}_{\ell_2} G^*(\bl_2',\bl_3)\tilde{C}_{\ell_2'} +{\rm 35\;perm.} + \nonumber \\
& &(2\pi)^6\delta^{(2)}(\bl_1-\bl_1')\delta^{(2)}(\sum\limits_i \bl_i)\delta^{(2)}(\sum\limits_i \bl_i')C_{\ell_1} \int \mathrm{d}^2\bm{\hat{n}} C_{|\bl_2-\bl_2'|}^{\phi\phi} G(\bl_2,\bl_2')\tilde{C}_{\ell_2} G(\bl_3,\bl_3')\tilde{C}_{\ell_3} +{\rm 35\;perm.}
\ee
where we have defined 
\be
G(\bl_i,\bl_j) &=& \bl_i \cdot ( \bl_i - \bl_j) e^{i \bm{\hat{n}} \cdot (\bl_i-\bl_j)},
\ee
and used 
\be
\delta^{(2)}(\bl_i+\bl_j) = \int \frac{\mathrm{d}^2\bm{\hat{n}}}{(2\pi)^2} e^{i (\bl_i+\bl_j) \cdot \bm{\hat{n}}}.
\ee
The second and third term contain an explicit coupling which we rewrote in terms of an integral over the line of sight direction. This was done to highlight the similarity with the full-sky result in the next section. In the language of field theory these terms present 1-loop corrections. The first term does not have such a loop structure. Instead there is a simple coupling described by the inner product between $\bl_2$/$\bl_2'$ and $\bl_1$. We find that this term dominates the lensing induced contribution to the covariance. The 36 permutations of the above term are introduced by the symmetry breaking of the $2p\times4p$ structure. However, when we compute the variance in Sec. \ref{sec:sizeOfEffect}, symmetries in the multipoles are restored and these permutations introduce an overall multiplicity.
\end{widetext}

\subsection{Full-Sky Temperature} \label{sec:fullSkyTres}
In this section we derive the full-sky expression for the lensing induced bispectrum covariance. The main purpose of this calculation is to obtain a better estimate of the effect on large angular scale (i.e., $\ell \lesssim 10$). As discussed earlier, squeezed configurations are expected to be the most important and hence an accurate computation of the extra covariance on large scales should provide a much better estimate than the flat-sky approximation. 

Let us write the lensed $a_{\ell m}$ in terms of the potential and the unlensed $\tilde{a}_{\ell m}$
\be
a_{\ell m} &=& \tilde{a}_{\ell m} + \int \mathrm{d}^2\bm{\hat{n}} Y_{\ell m}^*(\bm{\hat{n}}) \nabla_i \phi(\bm{\hat{n}}) \nabla^i \tilde{T}(\bm{\hat{n}}) + ...,
\ee
where $\tilde{T}(\bm{\hat{n}})$ is the unlensed CMB temperature in direction $\hat{n}$. We can expand the second term in spherical harmonics as well (working to lowest order in the lensing potential $\phi$) 
\be
a_{\ell m} &=& \tilde{a}_{\ell m} + \sum\limits_{LM} \sum\limits_{\ell'm'}\phi^*_{LM} \tilde{a}^{*}_{\ell'm'} \left(\begin{matrix} \ell & L & \ell' \\ m & M & m' \end{matrix} \right)F_{\ell L \ell'}.\nonumber \\
\ee
Here the coupling matrix is given by \cite{Hu2001}
\be
F_{\ell L \ell'} &=&\left[\frac{(2\ell+1)(2\ell'+1)(2L+1)}{4\pi}\right]^{1/2} \nonumber \\
&&\sqrt{L(L+1)\ell'(\ell'+1)} \left( \begin{matrix} \ell & \ell' & L \\ 0 & -1 & 1 \end{matrix} \right),
\ee
which is non-zero only when $\ell+\ell'+L = $ even. 

We aim to compute
\begin{widetext}
\be
{\rm Var}(\widehat{B}) \equiv \langle \widehat{B}_{\ell_1 \ell_2 \ell_3} \widehat{B}^*_{\ell_1' \ell_2' \ell_3'}\rangle &=&\sum\limits_{mm'} \left( \begin{matrix} \ell_1 & \ell_2 & \ell_3 \\ m_1 & m_2 & m_3 \end{matrix} \right) \left( \begin{matrix} \ell_1' & \ell_2' & \ell_3' \\ m_1' & m_2' & m_3'\end{matrix} \right)\langle a_{\ell_1 m_1} a_{\ell_2 m_2} a_{\ell_3 m_3} a^*_{\ell_1' m_1'} a^*_{\ell_2' m_2'} a^*_{\ell_3' m_3'}\rangle. 
\ee
For simplicity we define
\be
Q_{\ell m} \equiv & \sum\limits_{LM} \sum\limits_{\ell'm'}\phi^*_{LM} \tilde{a}^{*}_{\ell'm'}\left(\begin{matrix} \ell & L & \ell' \\ m & M & m' \end{matrix} \right)F_{\ell L \ell'}.
\ee 
We can then write schematically 
\be
\Delta{\rm Var}(\widehat{B})_{2p\times4p} &=& \sum\limits_{mm'} \left( \begin{matrix} \ell_1 & \ell_2 & \ell_3 \\ m_1 & m_2 & m_3 \end{matrix} \right) \left( \begin{matrix} \ell_1' & \ell_2' & \ell_3' \\ m_1' & m_2' & m_3'\end{matrix} \right) \delta_{\ell_1 \ell_1'} \delta_{m_1 m_1'} C_{\ell_1} \left[\overbrace{\langle \tilde{a}_{\ell_2 m_2}\tilde{a}_{\ell_3 m_3} Q_{\ell_2'm_2'}^*Q^*_{\ell_3'm_3'}\rangle_{\rm c}}^{(1)} + \right. \nonumber \\
&& \left. \overbrace{\langle \tilde{a}_{\ell_2m_2}Q_{\ell_3m_3}\tilde{a}^{*}_{\ell_2'm_2'}Q^*_{\ell_3' m_3'}\rangle_{\rm c}}^{(2)} \right]
 + {\rm perm}+\mathcal{O}(\phi^3)
\label{eq:BisCovFullSky}
\ee
where the subscript c shows that we keep only the connected terms. We compute these contributions term by term and the details can be found in Appendix \ref{app:fullSkyderivation}. After some algebra we find
\be
\Delta{\rm Var}(\widehat{B})_{2p\times4p} &=& \delta_{\ell_1\ell_1'}C_{\ell_1} C_{\ell_1}^{\phi\phi}F_{\ell_3 \ell_1 \ell_2}\tilde{C}_{\ell_2} F_{\ell_3' \ell_1 \ell_2'}\tilde{C}_{\ell_2'} 
\frac{1}{2\ell_1+1}+{\rm 35\;perm.} +\nonumber \\
&& \delta_{\ell_1\ell_1'}C_{\ell_1}(-1)^{\ell_3+\ell_2'}\sum\limits_L C_L^{\phi\phi}\left\{ \begin{matrix} \ell_1 & \ell_3 & \ell_2 \\ L & \ell_2' & \ell_3' \end{matrix} \right\}F_{\ell_2'L\ell_2} \tilde{C}_{\ell_2}F_{\ell_3'L\ell_3} \tilde{C}_{\ell_3} + {\rm 35\; perm.} +\nonumber \\
&&\delta_{\ell_1\ell_1'}C_{\ell_1} (-1)^{\ell_2+\ell_2'}\sum\limits_L C_L^{\phi\phi}\left\{ \begin{matrix} \ell_1 & \ell_2 & \ell_3 \\ L & \ell_2' & \ell_3' \end{matrix} \right\}F_{\ell_3'L\ell_2} \tilde{C}_{\ell_2} F_{\ell_3L\ell_2'}\tilde{C}_{\ell_2'} + {\rm 35\; perm.}
\label{eq:BisCovFullSkyEval}
\ee
\end{widetext}
The above has an identical structure to the flat-sky result. The are three distinguishable contributions. The last two contain an explicit internal sum with a coupling described by a Wigner 6-$j$ symbol (see Appendix \ref{app:fullSkyderivation}). In this form we can see the correspondence to terms found in studying the CMB lensing power spectrum. The first term is analogous to the reconstructed $C^{\phi \phi}$ term whilst the other terms are analogous to the N1 bias terms \citep{Kesden2003,Hanson2011}. Note that the normal Gaussian term would be analogous to the N0 bias. The second and the third contribution are challenging to compute numerically because the Wigner 6-$j$ is slow to compute. Fortunately, the first term dominates this sum and so in this work we focus on computing the first term (this assumption is verified in Appendix \ref{app:N1check}). We note that the first term still has five independent degrees of freedom, which equates to a five dimensional sum when estimating Var$(f_{\rm NL})$ and so is slow to evaluate. 

\subsection{Generalisation to polarisation}
For simplicity in the previous section we focused on the effect on temperature only maps, however non-Gaussianity estimators use $E$ and $B$ mode maps as well and thus we need to include the effects of lensing on these fields.
\subsubsection{Flat Sky} 
First we perform the calculation in the flat sky. The bispectrum variance is now given by
\begin{widetext}
\be
\langle B^{X_1,X_2,X_3}{B^{X_1',X_2',X_3'}}^* \rangle = (2\pi)^2\delta^{(2)}(\sum\limits_i \bl_i) (2\pi)^2\delta^{(2)}(\sum\limits_i\bl_i') \langle {X_1}(\bl_1){X_2}(\bl_2){X_3}(\bl_3){{X_1'}}^*(\bl_1'){{X_2'}}^*(\bl_2'){{X_3'}}^*(\bl_3')\rangle. \nonumber \\
\ee
The effect of lensing on polarization maps is given by
\be
E(\bl) \pm i B(\bl) = \tilde{E}(\bl)\pm i \tilde{B}(\bl)-\dtl{}e^{\pm 2i(\psi_{\bl'}-\psi_{\bl})}\left(\tilde{E}(\bl')\pm i \tilde{B}(\bl') \right)\phi(\bl-\bl'_1)(\bl-\bl'_1)\cdot\bl'_1,
\ee
where $\psi_{\bl}$ is the angle between the vector $\bl$ and the $\ell_x$ axis.
Thus we see that the flat-sky results can be immediately generalized by making the following replacement\be
(\bl_1+\bl_2) \cdot \bl_2 \tilde{C}_{\ell_2} \leftrightarrow G^{X_1,X_2}_{\bl_1,\bl_2},
\ee
where we have defined $G^{X_1,X_2}_{\bl_1,\bl_2}$ as 
\be
G^{X_1,X_2}_{\bl_1,\bl_2} = \begin{cases} (\bl_1+\bl_2) \cdot \bl_2 \tilde{C}^{TX_2}_{\ell_2} , & \text{if      $X_1=T$}\\
								 (\bl_1+\bl_2) \cdot \bl_2 \left[\cos{2 (\psi_{\bl_1}+\psi_{\bl_2})}\tilde{C}^{EX_2}_{\ell_2}- \sin{2 (\psi_{\bl_1}+\psi_{\bl_2})} \tilde{C}^{BX_2}_{\ell_2} \right], & \text{if      $X_1=E$}\\
								 (\bl_1+\bl_2) \cdot \bl_2 \left[\cos{2 (\psi_{\bl_1}+\psi_{\bl_2})} \tilde{C}^{BX_2}_{\ell_2}+\sin{2 (\psi_{\bl_1}+\psi_{\bl_2})} \tilde{C}^{EX_2}_{\ell_2} \right], & \text{if      $X_1=B$}\\
						\end{cases}. \nonumber \\
\ee
Thus the first line of Eq. \eqref{eq:flatSky2x4pnt}, which is the leading contribution, is given by
\be
{\rm Var}(\widehat{B})_{2p\times4p} &=& (2\pi)^8\delta(\bl_1-\bl_1')\delta^{(2)}(\sum\limits_i \bl_i)\delta^{(2)}(\sum\limits_i \bl_i') C^{T,X_1,X_1'}_{\ell_1} C_{\ell_1}^{\phi\phi} G^{X_2,X_3}_{\bl_2,\bl_3} G^{X_2',X_3'}_{\bl_2',\bl_3'} +{\rm 35\;perm.}
\ee
\subsubsection{Full Sky}
Now working in the full sky, the bispectrum variance is given by
\be
\langle B^{X_1,X_2,X_3}_{\ell_1 \ell_2 \ell_3} {B^{X_1',X_2',X_3'}}^*_{\ell_1' \ell_2' \ell_3'}\rangle &=&\sum\limits_{mm'} \left( \begin{matrix} \ell_1 & \ell_2 & \ell_3 \\ m_1 & m_2 & m_3 \end{matrix} \right) \left( \begin{matrix} \ell_1' & \ell_2' & \ell_3' \\ m_1' & m_2' & m_3'\end{matrix} \right) 
 \langle a^{X_1}_{\ell_1 m_1} a^{X_2}_{\ell_2 m_2} a^{X_3}_{\ell_3 m_3} {a^{X_1'}}^*_{\ell_1' m_1'} {a^{X_2'}}^*_{\ell_2' m_2'} {a^{X_3'}}^*_{\ell_3' m_3'}\rangle. \nonumber \\ 
\ee

The effect of lensing on $E$ and $B$ modes (in the full sky) is 
\be
\delta a^E = \sum \limits_{L,M}\sum \limits_{\ell',m'} \left( \begin{matrix} \ell & L & \ell' \\ m & M & m' \end{matrix} \right) \phi^*_{L,M}\left[ F^{+2}_{\ell,L,\ell'} {\tilde{a}^{E^*}}_{\ell',m'}-i F^{-2}_{\ell,L,\ell'}{\tilde{a}^{B^*}}_{\ell',m'} \right] \nonumber \\
\delta a^B = \sum \limits_{L,M}\sum \limits_{\ell',m'} \left( \begin{matrix} \ell & L & \ell' \\ m & M & m' \end{matrix} \right) \phi^*_{L,M}\left[ F^{+2}_{\ell,L,\ell'} {\tilde{a}^{B^*}}_{\ell',m'}+i F^{-2}_{\ell,L,\ell'}{\tilde{a}^{E^*}}_{\ell',m'}\right],
\ee
where 
\be
F^{\pm s}_{\ell,L,\ell'} = \frac{1}{4} \left[ L(L+1)+\ell'(\ell'+1)-\ell'(\ell'+1) \right] \sqrt{\frac{\ell(\ell+1)L(L+1)\ell'(\ell'+1) }{4\pi}}\left[ \left( \begin{matrix} \ell & L & \ell' \\ s & 0 & -s \end{matrix} \right) \pm \left( \begin{matrix} \ell & L & \ell' \\ -s & 0 & +s'\end{matrix} \right) \right]. 
\ee
\end{widetext}
We note that $F^{+s}_{\ell,L,\ell'}$ is only non zero when $\ell+L+\ell' =$ even and that $F^{-s}_{\ell,L,\ell'}$ is only non zero when $\ell+L+\ell' =$ odd. As the manipulations in Section \ref{sec:fullSkyTres} did not utilize any properties of the lensing kernel $F_{\ell,L,\ell'}$ it is trivial to generalise to the polarisation result, with the caveat that not all of the permutations are equivalent. To obtain the general result we perform the following replacement
\be
F_{\ell_1L\ell_2} \tilde{C}_{\ell_2} \leftrightarrow G^{X_1,X_2}_{\ell_1 L\ell_2},
\ee
where we have defined $G^{X_1,X_2}_{\ell_1L\ell_2}$ as 
\be
G^{X_1,X_2}_{\ell_1L\ell_2} = \begin{cases} F_{\ell_1L\ell_2} \tilde{C}^{TX_2}_{\ell_2} , & \text{if      $X_1=T$}\\
								 F^{+2}_{\ell_1L\ell_2}\tilde{C}^{EX_2}_{\ell_2}-i F^{-2}_{\ell_1L\ell_2} \tilde{C}^{BX_2}_{\ell_2} , & \text{if      $X_1=E$}\\
								 F^{+2}_{\ell_1L\ell_2} \tilde{C}^{BX_2}_{\ell_2}+i F^{-2}_{\ell_1L\ell_2} \tilde{C}^{EX_2}_{\ell_2} , & \text{if      $X_1=B$}\\
						\end{cases}. \nonumber \\
\ee
As a concrete example, the first term in Eq.~\eqref{eq:BisCovFullSkyEval} would be 
\be \label{eq:polarizedExample}
\Delta{\rm Var}(\widehat{B})_{2p\times4p} & = & \delta_{\ell_1\ell_1'}C^{T,X_1,X_1'}_{\ell_1} C_{\ell_1}^{\phi\phi}\frac{G^{X_2,X_3}_{\ell_3 \ell_1 \ell_2}G^{X_2',X_3'}_{\ell_3' \ell_1 \ell_2'}}{2\ell_1+1}. \nonumber \\
\ee
Note that as $F^{-s}_{\ell,L,\ell'}$ is only non zero when $\ell+L+\ell' =$ odd, some of these terms simplify if the bispectrum estimator only measures even parity ($\ell+L+\ell' =$ even), or odd parity ($\ell+L+\ell' =$ odd) bispectra. Thus for scalar primordial bispectra, which typically have $\ell_1+\ell_2+\ell_3 =$ even, the contributions to \eqref{eq:polarizedExample} from $B$-mode fields will be zero. 

\section{Relation to Primordial non-Gaussianity constraints}\label{sec:fnlEstimators}

The above discussion has focused on the variance of the angle averaged bispectrum or the full-bispectrum in the flat sky. In practice, calculating either of these has some challenges. First, the expected signal-to-nose ratio (SNR) in any triplet of these estimators is expected to be small. Second, explicitly calculating all of the bispectrum triplets is computationally prohibitive. To overcome these issues it is common to measure template amplitudes instead defined as  \citep{Komatsu2005} 
\begin{widetext}
\be
\hat{f}_{\rm NL}^{i} = N \sum\limits_{\ell_i,X_i,Y_i}b^{i,X_1,X_2,X_3}_{\ell_1,\ell_2,\ell_3} \begin{pmatrix}\ell_1 & \ell_2 & \ell_3 \\ m_1 & m_2 & m_3 \end{pmatrix} W^{X_1,X_2,X_3,Y_1,Y_2,Y_3}_{\ell_1,m_1,\ell_2,m_2,\ell_3,m_3} a^{Y_1}_{\ell_1 m_1} a^{Y_2}_{\ell_2 m_2} a^{Y_3}_{ \ell_3 m_3},
\ee
where $\hat{f}_{\rm NL}^{i}$ is the estimated amplitude of the bispectrum with reduced bispectrum $ b^i(\ell_1,\ell_2,\ell_3) $, $W^{X_1,X_2,X_3,Y_1,Y_2,Y_3}_{\ell_1,m_1,\ell_2,m_2,\ell_3,m_3} $ is a set of weight functions and $N$ is the estimator normalization.  For the minimum variance estimator (in the case of weak primordial non-Gaussianity and Gaussian CMB) the weight functions are separable and correspond to inverse variance filtering.

The variance of the template amplitudes is given by

\be
\langle \hat{f}_{\rm NL}^{i}\hat{f}_{\rm NL}^{i} \rangle &=& N^2 \sum\limits_{\ell_i,X_i}\sum\limits_{\ell'_i,X'_i}  b^{i,X_1,X_2,X_3}_{\ell_1,\ell_2,\ell_3}\begin{pmatrix}\ell_1 & \ell_2 & \ell_3 \\ m_1 & m_2 & m_3 \end{pmatrix} \times \nonumber \\ && \langle  C^{-1}[a]^{X_1}_{\ell_1 m_1} C^{-1}[a]^{X_2}_{\ell_2 m_2} C^{-1}[a]^{X_3}_{\ell_3 m_3} C^{-1}[a^*]^{X'_1}_{\ell_1' m_1'} C^{-1}[a^*]^{X'_2}_{\ell_2' m_2'}C^{-1} [a^*]^{X'_3}_{\ell_3' m_3'} \rangle\begin{pmatrix}\ell_1' & \ell_2' & \ell_3 '\\ m_1' & m_2' & m_3'\end{pmatrix}  b^{i,X_1',X_2',X_3'}_{\ell_1',\ell_2,'\ell_3'} ,
\ee
where $C^{-1}[a]^{X}_{\ell,m} \equiv \sum \limits_{\ell',m',X'}{{C}^{-1}}^{X,X'}_{\ell,m,\ell',m'} a^{X'}_{\ell',m'} $. When the covariance is diagonal in $m$ space (which is generally a reasonable approximation), this can then be simply related to the results of the previous section as
\be
\langle \hat{f}_{\rm NL}^{i} \hat{f}_{\rm NL}^{i} \rangle &=& N^2 \sum\limits_{\ell_i,X_i} \sum\limits_{\ell'_i,X'_i}b^{i,X_1,X_2,X_3}_{\ell_1,\ell_2,\ell_3} C^{-1}C^{-1}C^{-1} \langle B^{X_1,X_2,X_3}_{\ell_1,\ell_2,\ell_3} B^{X_1',X_2',X_3'}_{\ell_1',\ell_2',\ell_3'}\rangle C^{-1}C^{-1}C^{-1} b^{i,X_1',X_2',X_3'}_{\ell_1',\ell_2,'\ell_3'},
\ee
where for clarity we suppressed the indices on the inverse covariance matrix. The template variance is thus just a weighted sum of the variance of the angle averaged bispectrum. For example we find the leading order temperature-only contribution to be
\be 
\langle \hat{f}_{\rm NL}^{i} \hat{f}_{\rm NL}^{i} \rangle_{2p\times 4p} &=& 36 N^2 \sum\limits_{\ell_i}\sum\limits_{\ell'_i}  b_{\ell_1,\ell_2,\ell_3} C^{-1}C^{-1}C^{-1} \delta_{\ell_1\ell_1'}C_{\ell_1} C_{\ell_1}^{\phi\phi}F_{\ell_3 \ell_1 \ell_2}\tilde{C}_{\ell_2} F_{\ell_3' \ell_1 \ell_2'}\tilde{C}_{\ell_2'} 
\frac{1}{2\ell_1+1} C^{-1}C^{-1}C^{-1} b_{\ell_1',\ell_2',\ell_3'}. 
\ee
In the flat-sky regime the sums are replaced with integrals, the $\{ \ell, m\}$ pairs with vectors $\bl$ and the Wigner 3-$j$ with dirac delta functions $\delta^{(2)}(\sum\limits_i \bl_i)$. For example the leading order term is given by
\be
\langle \hat{f}_{\rm NL}^{i} \hat{f}_{\rm NL}^{i} \rangle_{2p\times 4p} &= 36 N^2 \int \prod_i \frac{\mathrm{d}^2\bl_i}{(2\pi)^2} \frac{\mathrm{d}^2\bl_i'}{(2\pi)^2}(2\pi)^8\delta^{(2)}(\sum \bl_i) \delta^{(2)} (\sum \bl_i') b_{\ell_1,\ell_2,\ell_3} \delta^{(2)}(\bl_1-\bl_1')C^{-1}C^{-1}C^{-1} \times \nonumber \\ & C_{\ell_1} C_{\ell_1}^{\phi\phi} \bl_1\cdot \bl_2 \tilde{C}_{\ell_2} \bl_1\cdot \bl_2' \tilde{C}_{\ell_2'}C^{-1}C^{-1}C^{-1} b_{\ell_1',\ell_2',\ell_3'}.
\ee
The subsequent sections explore how the lensing contributions impact measurements of the primordial template amplitudes.
\end{widetext}
\section{How large is the effect?}\label{sec:sizeOfEffect}

\begin{figure*}
\subfloat[Temperature Only ]{
 \centering
  \includegraphics[width=.45\textwidth]{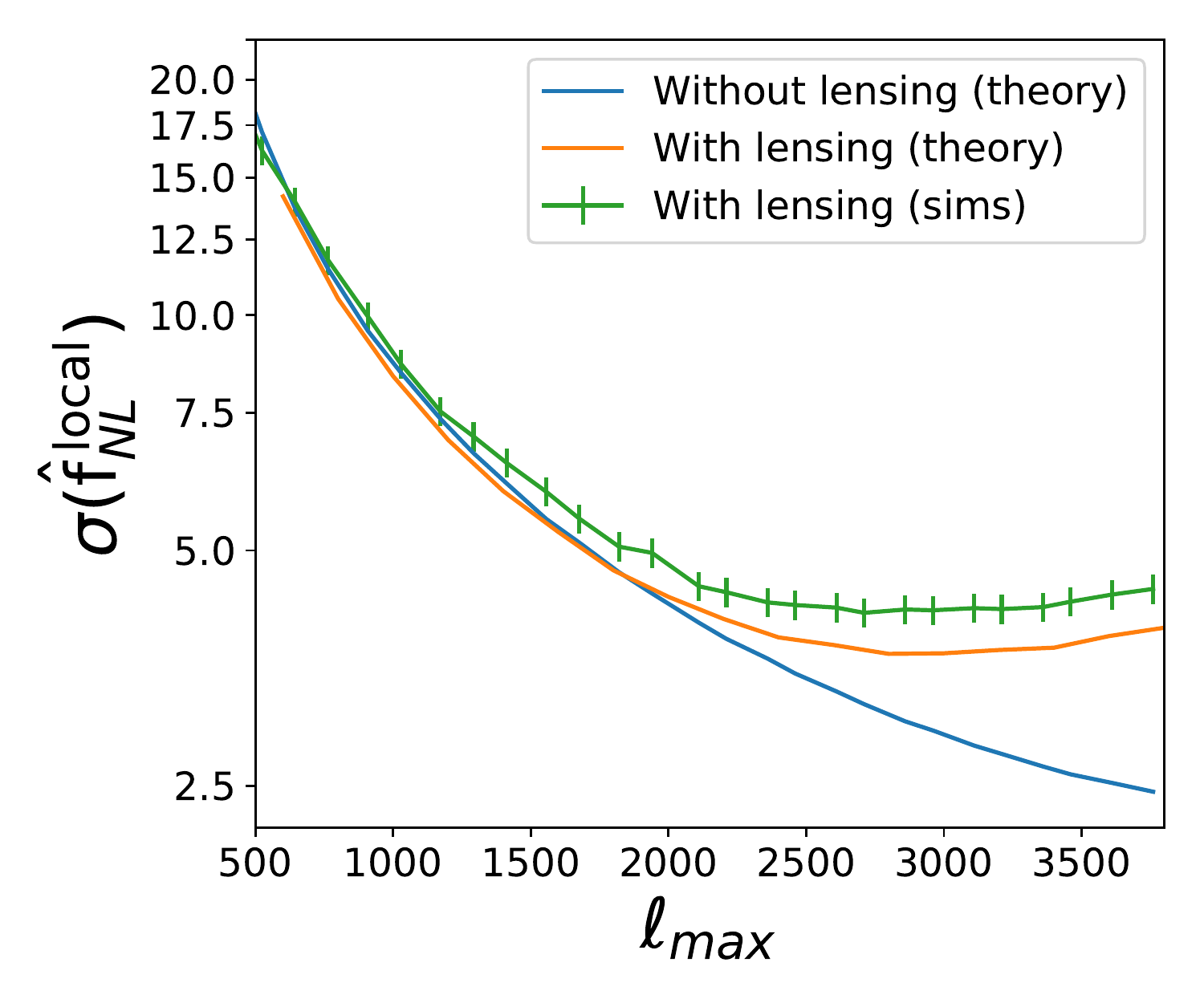}
  \label{fig:lensingVar_T}}
  \qquad
\subfloat[Temperature and $E$-mode polarization ]{
  \centering
  \includegraphics[width=.45\textwidth]{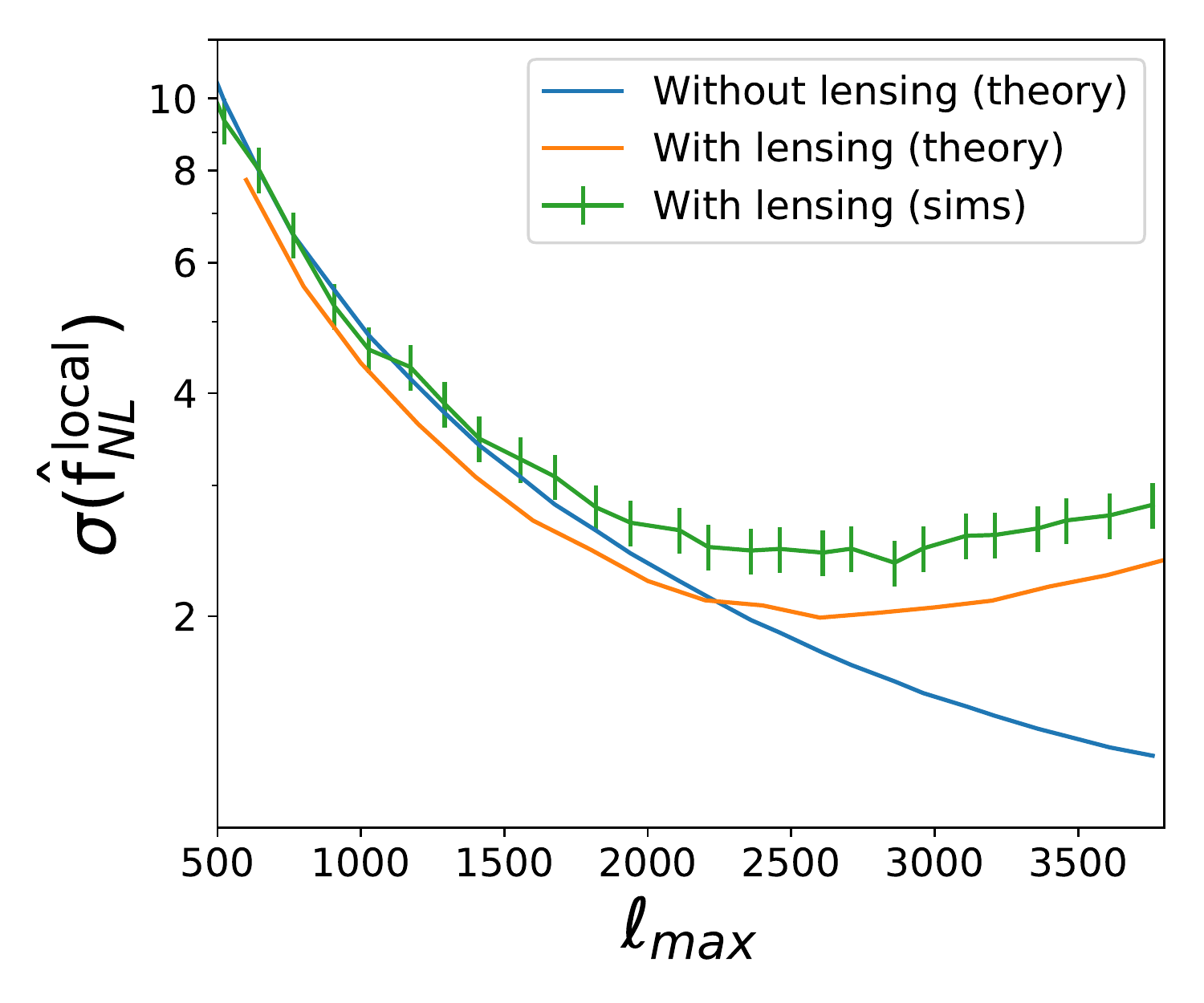}
  \label{fig:lensingVar_TE}}
\caption{The standard deviation of the estimator $\hat{f}^{\mathrm{local}}_{\mathrm{NL}}$ as a function of maximum scale, for the case of Gaussian CMB fields, blue curve, and including the lensing non-Gaussian contributions, orange curve. We also plot the measurements from our non-Gaussian simulations, green curve. We see that for $\ell_\mathrm{max}\gtrsim 2000$ the error begins to saturate due to the variance sourced by CMB lensing, which is the subject of this paper. We note that the Gaussian results include the effects of lensing in the power spectrum (i.e., we use the lensed $C_\ell$ to compute the standard deviation) but no higher order contributions. }
\end{figure*}

\begin{figure*}
\subfloat[Equilateral non-Gaussianity ]{
 \centering
  \includegraphics[width=.45\textwidth]{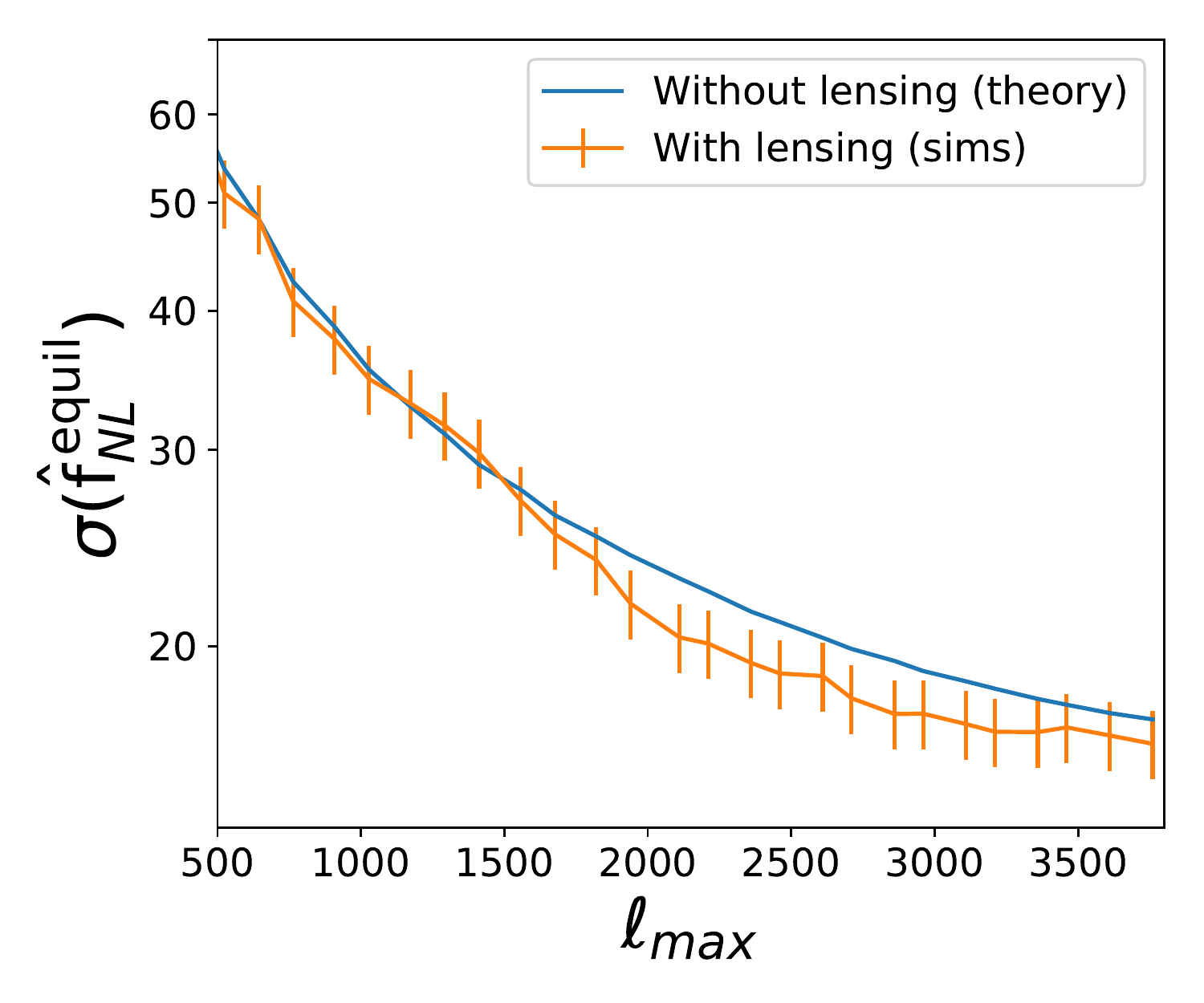}
  \label{fig:lensingVar_equil}}
  \qquad
\subfloat[Orthogonal non-Gaussianity ]{
  \centering
  \includegraphics[width=.45\textwidth]{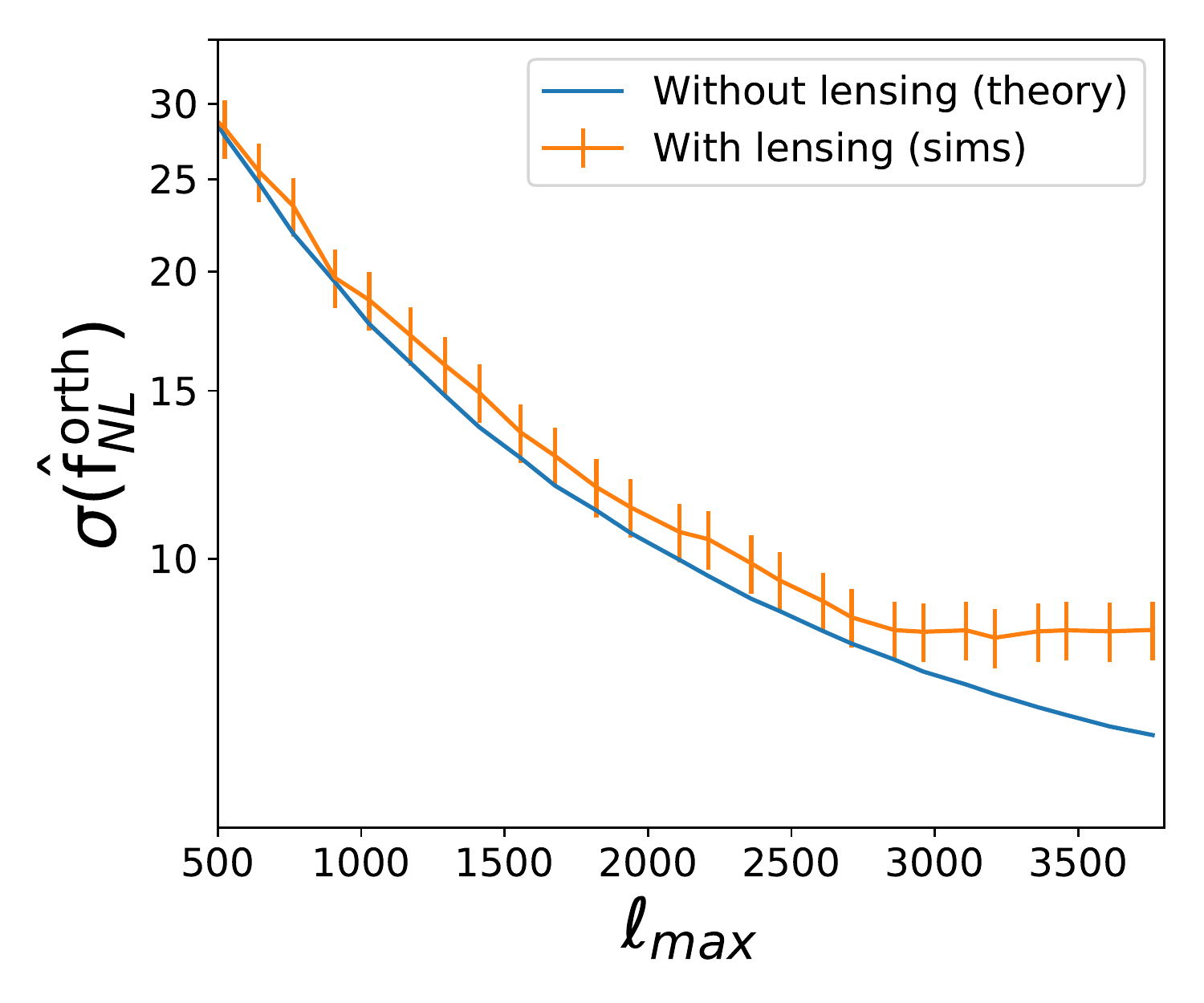}
  \label{fig:lensingVar_orth}}
\caption{The standard deviation of the estimators $\hat{f}^{\rm equil}_{\rm NL}$ and $\hat{f}^{\rm orth}_{\rm NL}$ for the case of Gaussian fields (without lensing) and including the lensing non-Gaussian contributions. Here we see that these configurations are not as affected by lensing as the local configuration. We again note that the Gaussian results includes the effects of lensing in the power spectrum (i.e., we use the lensed $C_{\ell}$ to compute the standard deviation) but no higher order contributions. }
\end{figure*}

\begin{figure*}
\subfloat[Idealized \textit{Planck} satellite ]{
 \centering
  \includegraphics[width=.45\textwidth]{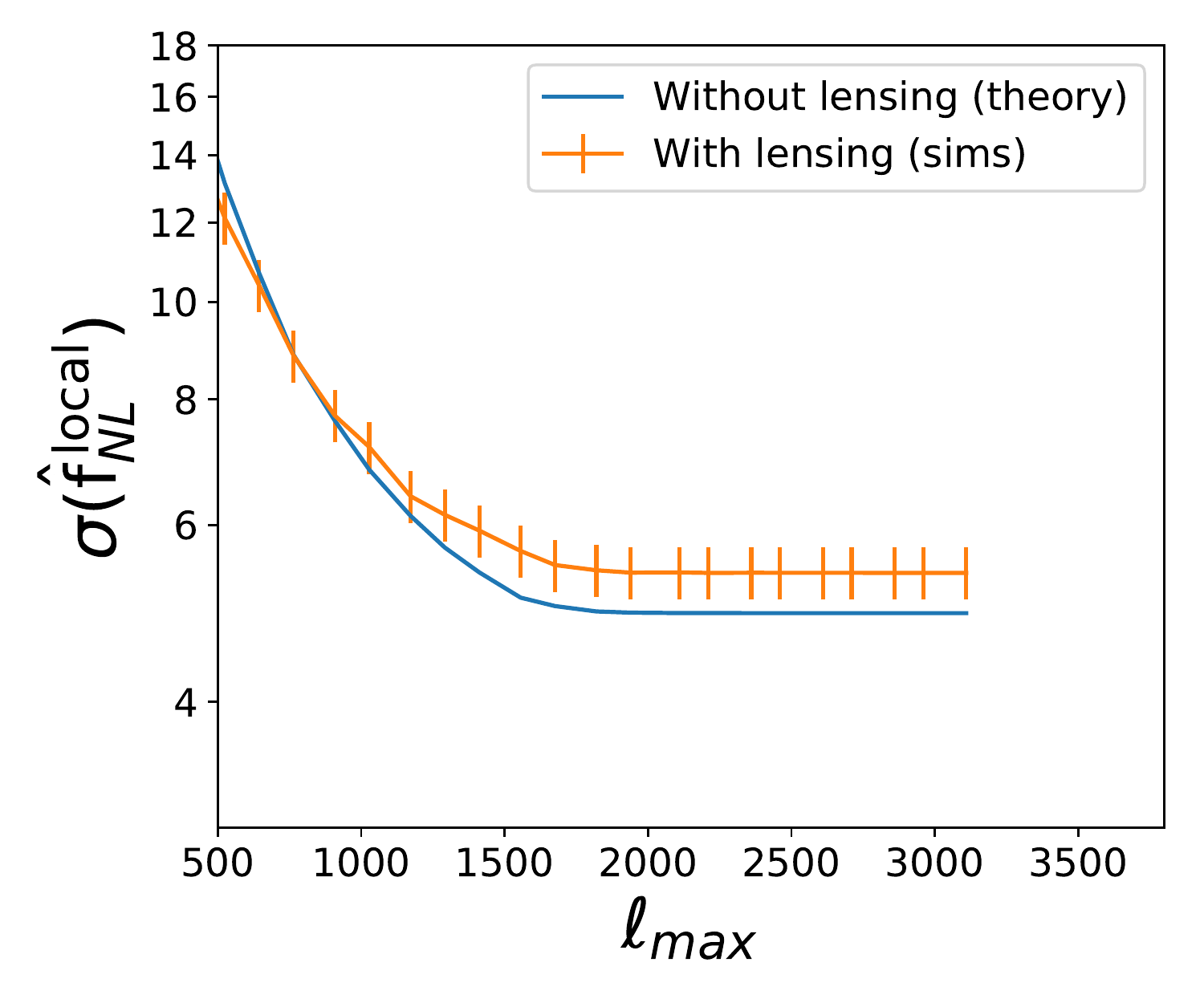}
  \label{fig:lensingVar_planck}}
  \qquad
\subfloat[Idealized Simons Observatory ]{
  \centering
  \includegraphics[width=.45\textwidth]{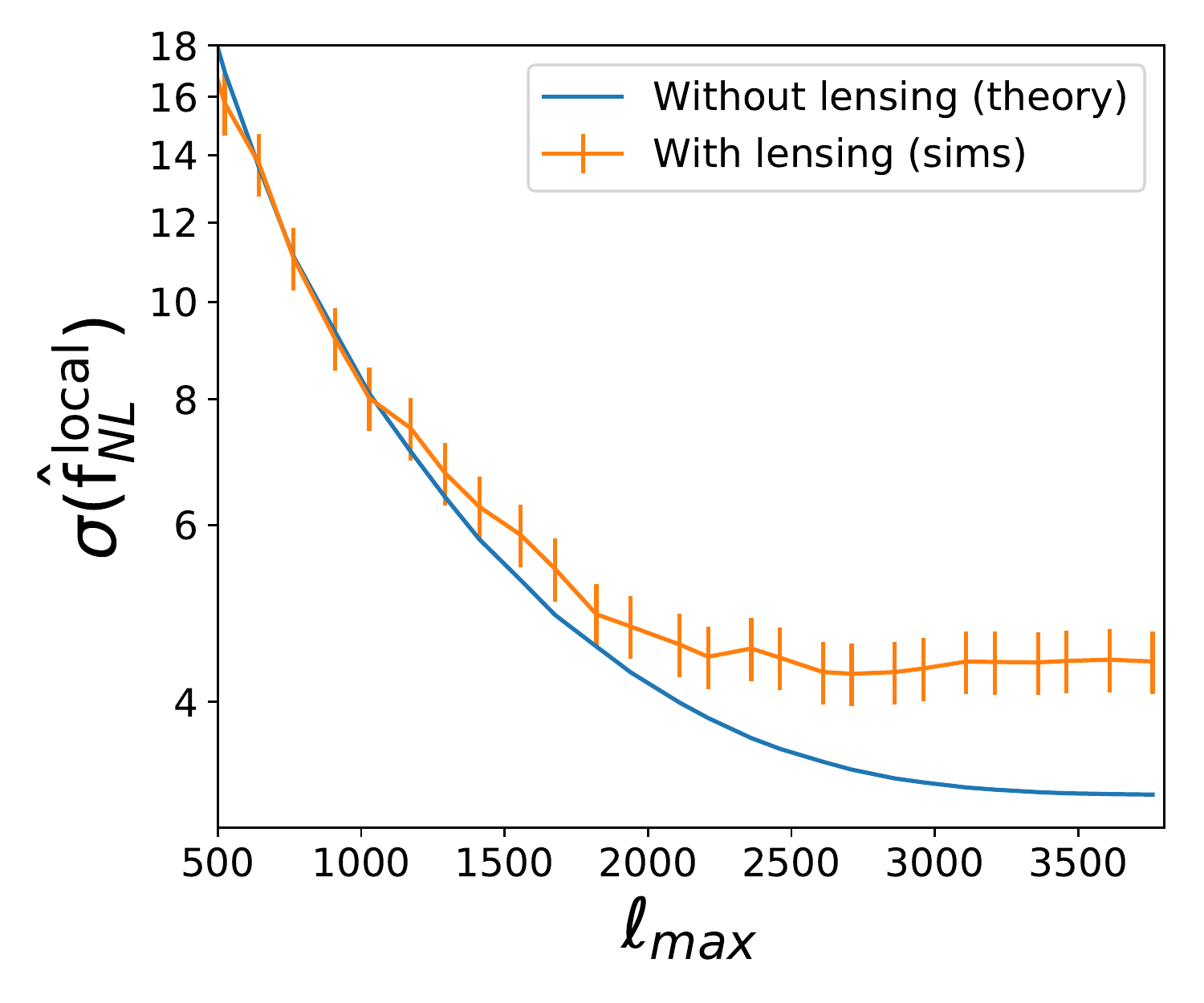}
  \label{fig:lensingVar_so}}
\caption{The standard deviation of the estimator $\hat{f}^{\rm local}_{\rm NL}$, for a \textit{Planck} like experiment and a Simons Observatory like experiment. The difference between these and Figs.~\ref{fig:lensingVar_TE} and \ref{fig:lensingVar_orth} is the inclusion of instrumental noise and scaling them to the observed sky fractions. This shows that the extra lensing variance, if unmitigated, would remove most of the improvements expected from an SO like experiment.}
\end{figure*}

We use the results of Section \ref{sec:effectOfLensing} to compute the size of the lensing induced variance on primordial local non-Gaussianity to first order in $C^{\phi \phi}$ for two cases: temperature only and temperature and $E$-mode polarization. Summing over all configurations that contribute to the six point function is computationally prohibitive as there are five or six nested sums that need to be evaluated to a large $\ell_{\rm max}$ as well as Wigner 6-$j$ terms, that are slow to calculate. In order to estimate the size of this term we only evaluate the leading term, shown in Eq. \eqref{eq:DB4}. We expect the term in Eq. \eqref{eq:DB4} will be the dominant contribution to the bispectrum variance and we checked this for low $\ell$, where the computation of all the terms is feasible, and found that to be the case. This is also justified from considering lensing reconstruction. The term in Eq. \eqref{eq:DB4}, which is analogous to the reconstructed $C^{\phi\phi}$ term, is the leading term and the  other terms, which are analogous to the N1 bias terms \citep{Kesden2003,Hanson2011}, are subdominant. Finally we checked this in simulations by isolating the non-leading terms, using the procedure described in Appendix \ref{app:N1check}, and found they contribute only around $\sim10\%$ of the total lensing variance. 

To further speed up the calculation, we used a combination of the full-sky and flat-sky expressions to compute the effect. We used the full-sky equations when $\ell_1$ in Eq.~\eqref{eq:BisCovFullSkyEval} satisfied $\ell_1 \le 40$ and the flat-sky equations for the rest. This enabled us to utilize the fast evaluation of the flat-sky terms. This also provides a robustness check on our code as we computed the suppression using the flat-sky and full-sky equations for a subset of configurations, those with $40 \le \ell_{1}<80$, and found percent level agreement.
Finally, we also replace the unlensed power spectra in Eq. \eqref{eq:DB4} with the lensed power spectra. In \cite{Lewis2011} they found that replacing the unlensed spectra with the lensed equivalents was a reasonable approximation to including higher order terms in $C^{\phi\phi}$ and we assume that this approximation will also improve our estimation of the extra variance.

In Figures \ref{fig:lensingVar_T} and \ref{fig:lensingVar_TE} we plot the bispectrum estimator standard deviation as a function of $\ell_{\rm max}$ for the temperature only and combined temperature and polarization cases. We find that this lensing trispectrum term becomes the dominant source of noise at $\ell>2000$ and means that pushing to scales much smaller than those modes already probed by \emph{Planck} results in no improvement to the local non-Gaussianity constraint. In Figures \ref{fig:lensingVar_T} and \ref{fig:lensingVar_TE} we also plot the results from analyzing a set of 553 temperature and 200 temperature and polarization lensed CMB realizations with a binned bispectrum estimator \citep{Bucher2010,Bucher2016} as implemented in \cite{Coulton2019}. The lensed simulations were generated with the public code pixell\footnote{https://github.com/simonsobs/pixell} which obtains lensed CMB maps on a cylindrical projection of the sky using bicubic spline interpolation. We find that the effect on the lensed simulation is similar to our first order calculation, despite the fact that the simulations are correct to all orders in $\phi$ (assuming that $\phi$ is Gaussian). Three points to note: first, the adjacent points from the binned bispectrum estimator are relatively strongly correlated. Second for small $\ell_{\rm max}$ we see that the theory prediction is less than the result without lensing. This is because our non-Gaussian theory calculation does not include the slight information loss that occurs due to our slightly suboptimal bin choice for our binned bispectrum estimator. Third,  at small scales we actually see an increase in the variance as we push to smaller scales, indicating a reduced ability to constrain non-Gaussianity. This highlights the sub-optimality of the standard estimator when the lensing variance contributions are large. We also verified that we obtain the same results, but without the degradation from binning, from using a KSW estimator\citep{Komatsu2005,Yadav2007}. These results are broadly consistent with the results of Ref.~\cite{Babich2004b}. We see a slightly smaller increase in the standard deviation at low $\ell$ and a slightly lower increase at high $\ell$, we find an increase in the standard deviation by $\sim 25\%$ at $\ell=2500$ whereas Ref.~\cite{Babich2004b} finds $\sim 30\%$. The differences are likely arise due to our treatment with the full-sky formalism, whereas Ref.~\cite{Babich2004b} only used the flat-sky approximation. 

In Figures \ref{fig:lensingVar_equil} and \ref{fig:lensingVar_orth} we explore the impact of lensing variance on searches for equilateral and orthogonal non-Gaussianity, using temperature and polarisation maps. We find that the equilateral shape is almost completely unaffected whilst the orthogonal shape is suppressed above $\ell =2500$. Note that as the standard deviation for the orthogonal shape decreases more slowly with $\ell_{\rm max}$ than the local shape, the constraints for the orthogonal shape are less impacted than the local shape \citep{Bartolo2009,Smith2011}. There is a simple explanation for why the local shape is more affected than equilateral shape. The local estimator probes squeezed shapes as local non-Gaussianity is the modulation of small scale power by a long wavelength mode. The effect of lensing on the CMB is a modulation of small scale power by degree scale lenses. This means that the part of the local bispectrum estimator which probes small scales is essentially performing a suboptimal quadratic lensing reconstruction \citep[see, e.g., Ref.~][for details on lensing reconstruction]{Okamoto2003}. This suboptimal lensing reconstruction means that there is a contribution to the estimator variance from the lensing potential power spectrum. For the equilateral configuration the lensing reconstruction is highly suboptimal, as it is less sensitive to the modulation of small scales by the large scales, and hence is less degraded.

Next we examine how important this effect is for real experiments. In Figure \ref{fig:lensingVar_planck} we plot the size of this for an idealised version of the \textit{Planck} satellite by using noise levels from \cite{Planck2006}. For simplicity we simulate full-sky maps as this drastically reduces the time to estimate the variance from simulations (by removing the need to calculate the normalisation from simulations and eliminating the linear term from the estimator; see e.g. \cite{Creminelli2006,Babich2005} for a description of the linear term and \cite{Smith2011} for a description of how this can be efficiently calculated with simulations). We approximate the effect of observing part of the sky by increasing the standard deviation by $1/\sqrt{f_{\rm sky}}$, where $f_{\rm sky}$ is the fraction of sky the experiment would observe and $f_{\rm sky}\approx 0.8$ for \textit{Planck} \citep{BenoitLevy2012,Schmittfull2013}. We find that for the \textit{Planck} experiment the lensing variance is a small effect, resulting in $<10\%$ increase in the noise. In Figure \ref{fig:lensingVar_so} we plot the size of this for an idealised version of the Simons Observatory (SO). The Simons Observatory is an upcoming CMB experiment located in the Atacama desert \citep{SO2019}. For simplicity we again use full sky maps with homogeneous noise to avoid requiring a linear term in our estimator. In reality the experiment will observe $\sim 40\%$ of the sky with anisotropic noise and we approximate these effects with the appropriate $f_{\rm sky}$ factor. We use noise power spectra, including the effects of atmosphere, from \cite{SO2019}. We find that SO is significantly affected by this effect with almost all the gains of pushing to smaller scales eradicated by the contribution of the lensing variance. 

There is a further potential, and slightly technical, complication caused by this extra lensing variance. When estimating non-Gaussianity from data it is important to accurately compute the estimator normalization, which is related to the six point function of the estimator (when applied to Gaussian maps). Typically the normalisation of the bispectrum estimator is calculated from simulations via a Monte Carlo average, for example as described in \cite{Smith2011}. If the maps used for this ensemble average include the effect of lensing, then there will be an additional, and unwanted, contribution to the estimator normalization from the lensing trispectum. This has the potential to introduce a bias that would reduce the non-Gaussianity signal (as the normalization would be too large). If bispectrum estimator normalisations are to be calculated from simulations then the simulations need to be Gaussian simulations.

\section{Removing the variance with ISW-lensing marginalization?} \label{sec:ISWmarg}

\begin{figure*}
  \centering
  \includegraphics[width=.45\textwidth]{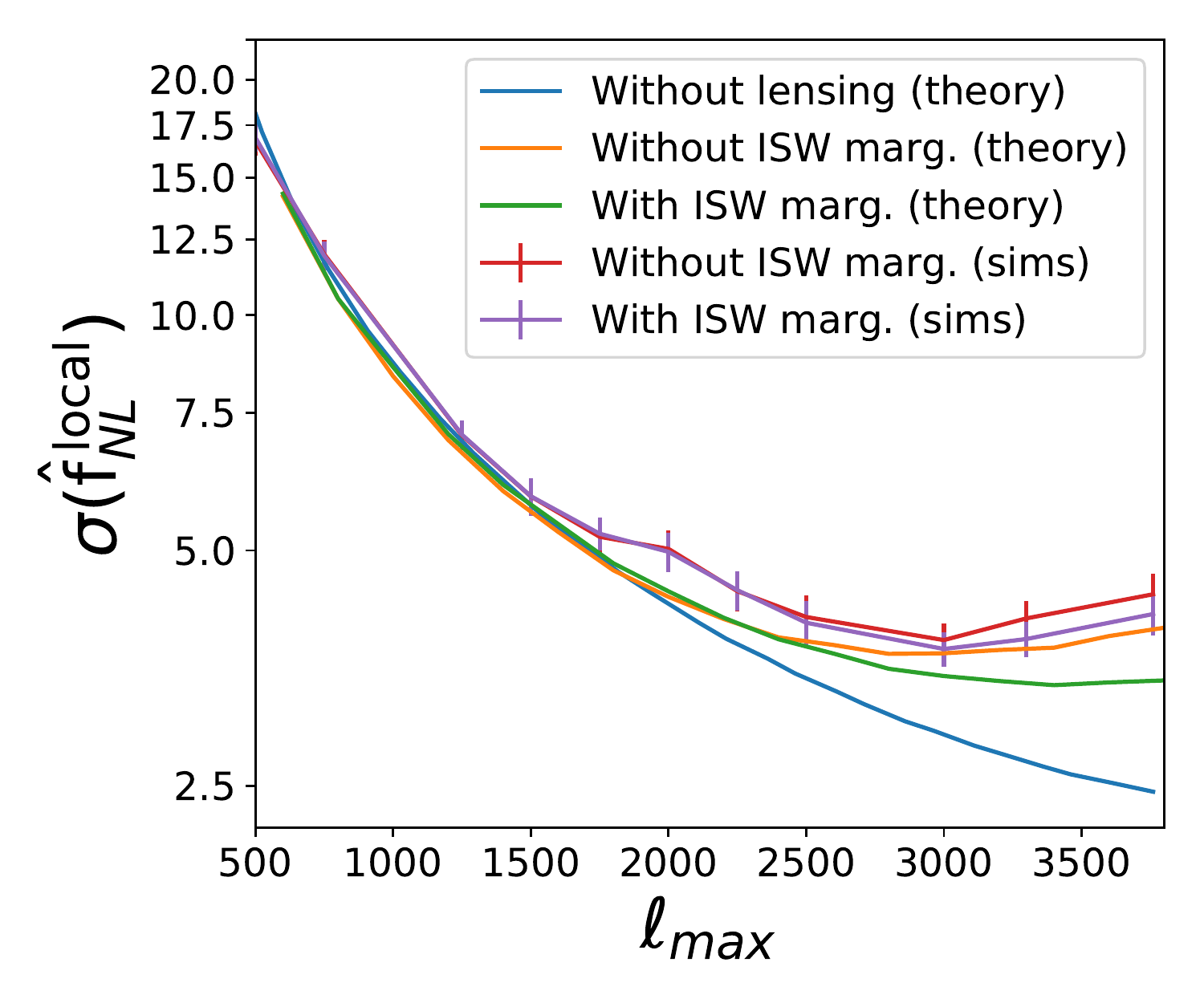}
\caption{ \label{fig:lensingVar_iswLensingMarg} The standard deviation of the estimator $\hat{f}^{\mathrm{local}}_{\mathrm{NL}}$, with ISW-lensing marginalisation. Theoretically, we find that ISW-marginalization reduces the contribution of the lensing 4-point function to the variance from the level predicted (orange) to the level shown in green. However when we compare the results from a suite of simulations with ISW-marginalization (purple) to a suite of simulations without (red) we see that the variance is only slightly reduced.}
\end{figure*}

One method for removing this variance would be to deproject the modes that are sourcing the extra variance from the estimator. Deprojecting modes will result in an increase of the standard deviation (as there are fewer modes to estimate primordial non-Gaussainity with), however a priori it is not obvious how much constraining power would be lost. This deprojection would be done by
\be
\hat{f}_{\rm NL}^{\mathrm{noise-free}} = \hat{f}_{\rm NL} - \sum \limits_{\ell_i,m_i} W_{\ell_1,\ell_2,\ell_3} a_{\ell_1,m_1} a_{\ell_2,m_2} a_{\ell_3,m_3} \nonumber \\
\ee
where the optimal weights depend on the full six point function. Computing the full six point function is complicated and it will be difficult to efficiently implement this optimally. However from the discussion of this effect in Section \ref{sec:sizeOfEffect} we know that the modes we wish to remove are actually of the following schematic form
\be
\hat{f}_{\rm NL}^{\mathrm{noise-free}} = \hat{f}_{\rm NL} - \sum \limits_{\ell_i,m_i} W_{\ell_1,\ell_2,\ell_3} a_{\ell_1,m_1} \hat{\phi}_{\ell_1,m_1}.
\ee
where $ \hat{\phi}$ is a quadratic reconstruction of the lensing potential. This form is very similar to a known bias to primordial non-Gaussianity searches, the ISW-lensing bispectrum \citep{[see e.g.][]Goldberg1999,Lewis2011}, and so suggests that this effect could be potentially reduced by marginalising over / jointly fitting for the amplitude of the ISW-lensing bispectrum.

Using a calculation to leading order in $C^{\phi\phi}_{\ell}$ we calculate the expected impact of ISW-lensing marginalisation, which is shown in Figure \ref{fig:lensingVar_iswLensingMarg}. We see that ISW-lensing marginalisation should be quite effective at removing this bias. In the same figure we also plot the result of using ISW-lensing marginalisation on a set of lensed simulations and we find that it is significantly less effective. Firstly we should note that this marginalisation is more effective than Figure \ref{fig:lensingVar_iswLensingMarg} implies. This is because performing the ISW-lensing marginalisation increase the estimator standard deviation (as it removes signal modes which are potentially contaminated) so the slight improvement seen in Figure \ref{fig:lensingVar_iswLensingMarg} means that the cost of performing the marginalisation is slightly outweighed by the removal of some of the lensing variance. Next we note that the ISW-lensing marginalisation is very sensitive to the shape of the measured bispectra. When the ISW-lensing marginalisation is performed using the binned estimator it is even less effective than the KSW estimator, whose results are plotted in Figure \ref{fig:lensingVar_iswLensingMarg}. This is because, unlike the KSW estimator, the binned estimator does not use the exact ISW-lensing template but rather a binned version of it for the ISW-lensing marginalisation. This result suggests an explanation for the difference between our theoretical calculation and the simulation results. It suggests that higher order terms distort the shape and reduce the effectiveness of this marginalization.

\section{Delensing}\label{sec:delensing}

\begin{figure*}
\subfloat[Local non-Gaussianity ]{
 \centering
  \includegraphics[width=.45\textwidth]{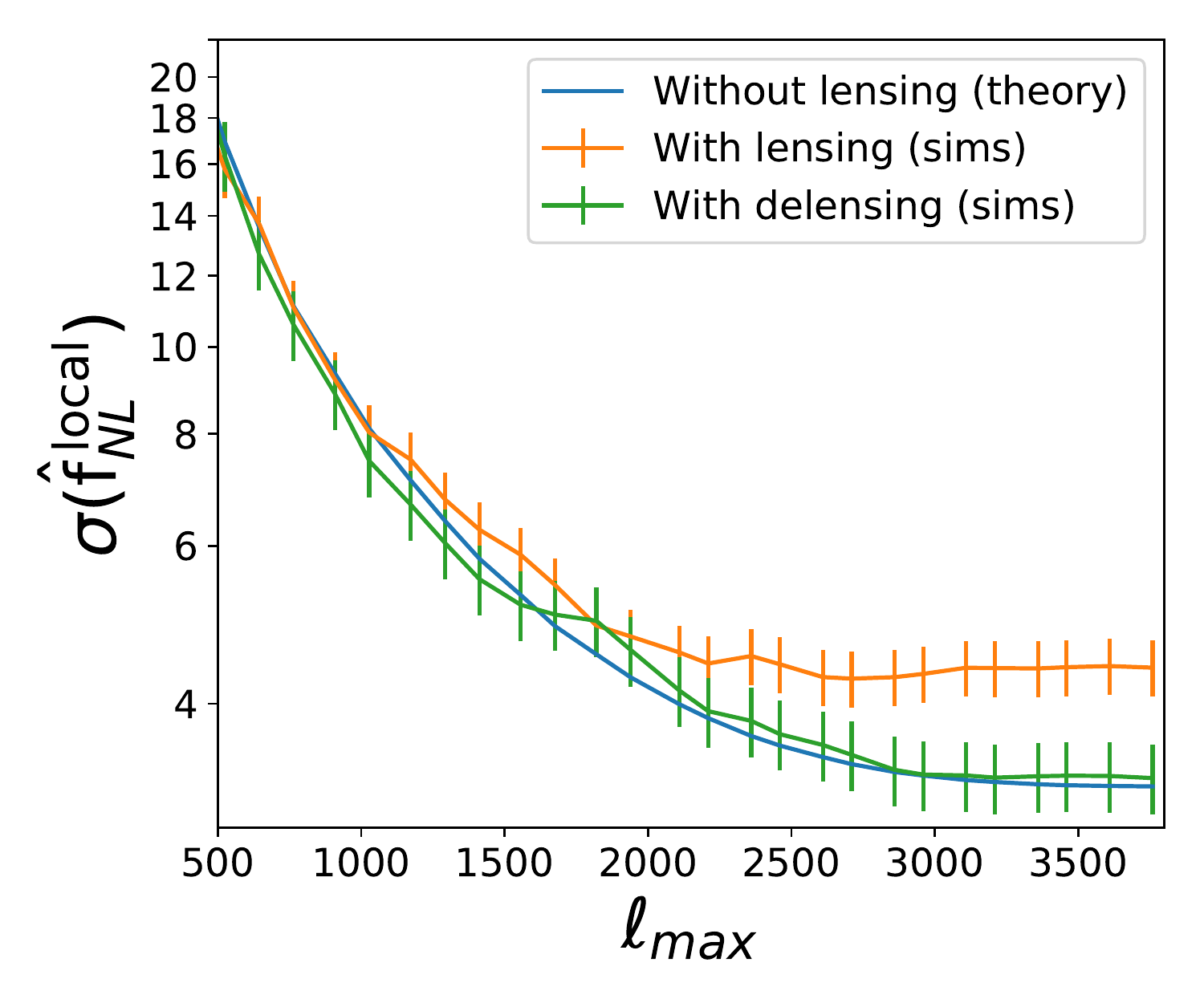}
  \label{fig:lensingVar_local_delens}}
  \qquad
\subfloat[Orthogonal non-Gaussianity ]{
  \centering
  \includegraphics[width=.45\textwidth]{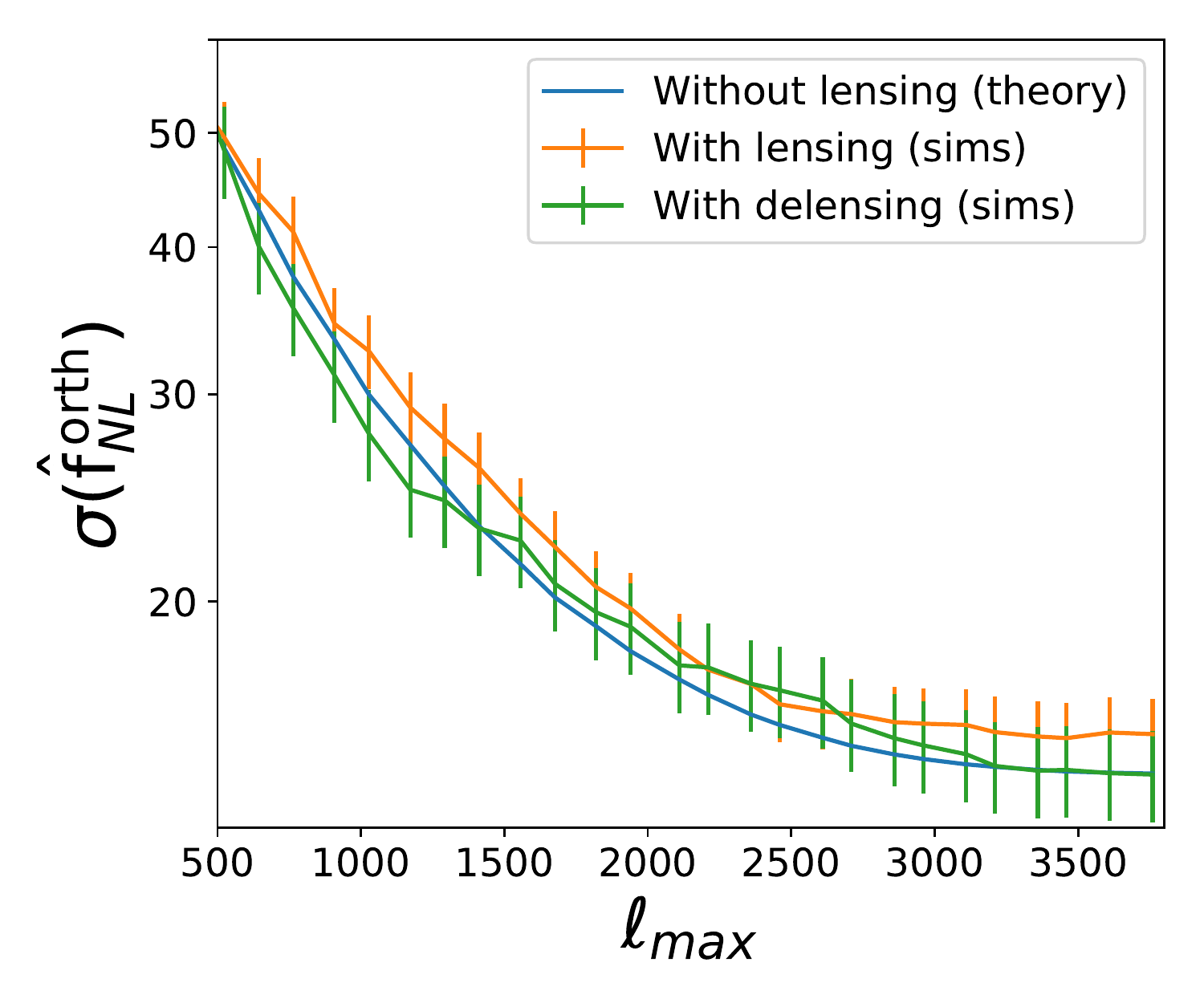}
  \label{fig:lensingVar_orth_delens}}
\caption{The standard deviation of the estimators $f^{local}_{NL}$ and $f^{orth}_{NL}$ after delensing for an idealized SO like experiment. As in Figure \ref{fig:lensingVar_TE} we show how the unlensed standard deviation (blue curve) is increased by the additional lensing induced variance (orange curve). However we find that, for an SO like experiment, delensing is able to almost completely remove the contribution of the lensing variance (green curve). This is the case for both the local and orthogonal shapes.}
\end{figure*}

Another possibility for removing this effect is delensing. Delensing is the process of remapping the observed temperature and polarisation measurements by an estimate of the lensing potential in order to estimate the unlensed fields \citep{Seljak2004,Smith2012}. Ref. ~\cite{Green2017} studied the impact of delensing on the lens-induced variance of the measured CMB power spectra, which is a four-point function over the CMB fields \cite{BenoitLevy2012}, and on the cross-covariance between the reconstructed lensing power spectrum and the CMB power spectra, which is a six-point function in terms of the CMB fields \cite{Schmittfull2013,Peloton2017}.  They found that delensing almost completely removes these lensing-induced variances, suggesting that a similar result might hold for the measurement of primordial bispectra. We note that the ISW marginalization discussed in Section \ref{sec:ISWmarg} is similar to the delensing procedure; the lensing field is reconstructed and subtracted off. However the weights are suboptimal (as modes are down-weighted by the ISW lensing power spectrum) and, even with optimal weighting, would only be equivalent to delensing at linear order; the delensing procedure used in this section captures part of the higher order terms.

Following the methods of \cite{Anderes2015, Larsen2016,Green2017} we use the measurements of the lensing potential to delens the CMB. The measured lensing potential could either come from reconstruction of the lensing potential from CMB maps \citep{Hu2001,Hu2002,Okamoto2003} or from using a tracer, such as the cosmic infrared background \citep{Simard2015,Sherwin2015}. For large scale lenses, which are the lenses we wish to remove, lensing reconstructions from CMB maps provide the highest fidelity lensing potential measurements. However this procedure can introduce biases from correlations between the lensing potential noise (which is related to the instrument noise and the CMB modes) and the temperature and polarization maps. When delensing the power spectrum, these biases can be significant \citep{Teng2011,Namikawa2014,Sehgal2017} and introduce additional non-Gaussianity \citep{Namikawa2015}. In this section we explore the effectiveness of using lensing potential measurements from CMB maps to remove the lensing contribution to the bispectrum variance. 

We use the minimum variance quadratic estimator \citep{Okamoto2003} to reconstruct the lensing potential using simulations of CMB maps as will be observed by SO. We use the public cmblensplus package\footnote{https://github.com/toshiyan/cmblensplus.git} to perform this reconstruction and follow \cite{Hanson2011} by using the lensed CMB fields in the estimator weights to reduce the higher order biases. We then use this reconstructed lensing potential map to delens the maps by remapping the pixels as
\be
\hat{\tilde{T}}(\mathbf{\hat{n}}) = \hat{T}(\mathbf{\hat{n}}-\nabla \hat{\phi}_{\rm Wiener}(\mathbf{\hat{n}})) \nonumber \\
\hat{\tilde{P}}(\mathbf{\hat{n}}) = R\hat{P}(\mathbf{\hat{n}}-\nabla \hat{\phi}_{\rm Wiener}(\mathbf{\hat{n}}))
\ee
where $P$ is the polarisation tensor and $R$ is a polarization rotation matrix which is very nearly unity \citep{Challinor2002}. Motivated by the results of \cite{Green2017}, we use Wiener filtered lensing potential maps. We then use the binned bispectrum estimator to measure the variance of a set of 218 of delensed simulations.

In Figures \ref{fig:lensingVar_local_delens} and \ref{fig:lensingVar_orth_delens} we examine how delensing would impact measurements of local and orthogonal non-Gaussianity for an idealized SO-like experiment. We find that delensing almost completely removes the contribution of lensing to the variance. Further we find no biases or extra contributions to the variance from higher order terms in the zero-signal limit. The lack of higher order biases can be simply understood. In a perturbative manner the effect of delensing can be described as 
\be
\hat{\tilde{T}}(\mathbf{\hat{n}}) = \hat{T}(\mathbf{\hat{n}}) - \nabla \hat{T}(\mathbf{\hat{n}}) \nabla \hat{\phi}_{\rm Wiener}(\mathbf{\hat{n}}).
\ee
Thus in bispectra analyses this will introduce bispectra terms of the schematic form
\be
\langle a_{\ell_1,m_1}a_{\ell_2,m_2} a_{\ell_3,m_3}\hat{\phi}_{\ell_3',m_3'} \rangle
\ee
As $\hat{\phi}$ is proportional to the product of two maps, this term will vanish for Gaussian fields. This argument extends to higher orders in $\hat{\phi}$ and so we do not expect biases of this form.
 There is the potential that by remapping these pixels we will lose some of the original bispectrum signal. It was shown in \cite{Pearson2012,Hanson2009} that the effect of lensing has a negligible effect on the primordial bispectrum signal. When we delens we are performing an identical operation to the original lensing operation and thus we expect the impact on the primordial signal to also be minimal. However, the delensing case is potentially more complicated as the reconstructed lensing potential may contain some of the primordial signal (which is not the case when the field is lensed with the true lensing potential). As discussed above, lensing can be thought of as modulating the small scale power by a large scale lens. When we delens, we reconstruct the large scale modulation field using the small scales and then remove it. Local non-Gaussianity is also a modulation of small scales by a large scale mode. This means that the reconstruction of the large scale lens may also `reconstruct' the primordial large scale modulation field, which is essentially the large scale CMB mode. Thus when we delens we could also be removing the primordial non-Gaussianity. The projection of the primordial field onto the lensing potential reconstruction is expected to be small and thus should have a small effect on the primodial non-Gaussainity amplitude. \footnote{We thank Antony Lewis for pointing out this complication.}
We investigated this effect using the following process. First we generated maps containing local non-Gaussainity using the methods described in \cite{Liguori2003,Elsner2009}. Then we lens these simulations, add noise for an SO like experiment and then delens, as described above. We then analyse these simulations with our non-Gaussianity pipeline. As these simulations are computationally intensive to run, we use a sample variance cancellation technique to reduce the required number of simulations. We analyse the same CMB realisation with $f^\mathrm{NL}_\mathrm{local}\neq 0 $ and also with $f^\mathrm{NL}_\mathrm{local} = 0$ and then look at the difference in the measured values of $f_{\rm local}$. Through this difference we can cancel the noise from the primary CMB and instrument. We find this reduces the number of simulations by two orders of magnitude. When simulating maps with $f^\mathrm{local}_\mathrm{NG} = 10 $ we find that after delensing the average recovered value is $\hat f^\mathrm{NL}_\mathrm{local} = 8.98 \pm 0.16$. This shows that delensing is introducing a multiplicative bias of $10.1\%\pm1.6\%$, which we believe is by the above mechanism. This bias would need to be accounted for in future analyses but as it is small it does not invalidate the use of delensing. Equivalently this bias would mean that the error bars would be boosted by $\sim 10\%$. This bias can be modelled by considering the leading order effect of delensing. We provide the details of the calculation in Appendix \ref{app:delensingBias} and here we summarise the result. The measured amplitude after delensing is given by
\begin{widetext}
\begin{align}
\langle \hat{f}^{\rm delensed}_{\rm NL} \rangle = f_{\rm NL}\bigg( & 1-\frac{6}{N} \sum \tilde{B}^{X_1,X_2,X_3}_{\ell_1,\ell_2,\ell_3}C^{X_2,S_3}_{\ell_2} F^{S_3}_{\ell_3,\ell_1,\ell_2}W_{\ell_1}A_{\ell_1} B^{X_1,\alpha,\beta}_{\ell_1,\ell_a,\ell_b}{g^{\alpha,\beta}}^*_{\ell_a,\ell_b,\ell_1} \frac{1}{2\ell_1+1} \nonumber \\
&-\frac{6}{N} \sum \tilde{B}^{X_1,X_2,X_3}_{\ell_1,\ell_2,\ell_3} B^{X_1,X_2,\beta}_{\ell_1,\ell_2,\ell_3}C^{S_3,\alpha}_{\ell'} F^{S_3}_{\ell_3,L,\ell'}W_L A_{L} {g^{\alpha\beta} }^*_{\ell',\ell_3,L}\frac{1}{2\ell_3+1}\nonumber \\
& -\frac{6}{N} \sum \tilde{B}^{X_1,X_2,X_3}_{\ell_1,\ell_2,\ell_3} B^{X_1,S_3,\beta}_{\ell_1,\ell',\ell_b}C^{X_2,\alpha}_{\ell'} F^{S_3}_{\ell_3,L,\ell'} W_LA_{L} {g^{\alpha\beta}}^*_{\ell_2,\ell_b,L}\frac{(-1)^{\ell_3+L}}{2\ell_1+1}\left\{ \begin{matrix} \ell_3 & \ell_2 & \ell_1 \\ \ell_b &\ell' & L \end{matrix} \right\} \bigg)
\end{align}
\end{widetext}
where $W_LA_{L}$ is the Wiener filter used when delensing the maps, $ S_i=T$ if $X_i= T$ and $S_i \in [E,B]$ otherwise and ${g^{\alpha\beta}}_{\ell,\ell',L}$ is the lensing kernel used to reconstruct the maps with the indices $\alpha$ and $\beta$ labelling the maps used to reconstruct the lensing potential\citep[see e.g][for examples of how delensing has been applied to existing data]{Carron2017,Sehgal2017}. This result shows that the effect is a multiplicative bias on the measured amplitude. Computing this bias for the Simons Observatory setup used above, we find that the bias would be $11.5\%$, in agreement with the result from simulations. We note that the form of this bias suggests a simple method to avoid it. Scalar bispectra receive no contribution from B modes. If we only used combinations of fields involving B modes to reconstruct the lensing potential then there would be no bias contributions from the first term and significantly reduced contributions from the second term (as these terms will be suppressed by the bispectrum parity condtion). With our suite of non-Gaussian simulations we verified that there was no bias when delensing was performed with a lensing potential containing only combinations involving one B mode contribution. For experiments beyond SO, quadratic estimators for lensing reconstruction will gain most of their signal to noise from configurations involving B modes (particularly the EB pair) and thus this restriction can be enforced with only a marginal cost to the lensing reconstruction accuracy.

\section{Discussion and Conclusions}
Primordial non-Gaussianity remains one of most important targets in cosmology for the purpose of constraining the physics of the early Universe. The CMB has provided us with the strongest constraints so far. Future large scale structure measurements as well as innovative new estimators that rely on cosmic variance cancellation will help closing in on $f_{\rm NL} \sim 1$. Although well studied, there is still more we can learn from the CMB. Simple mode counting arguments suggest that improvement of a factor of $2-4$ are possible with next generation CMB experiments such as SO and CMB-S4. In this paper we addressed one major obstacle reaching those forecasted numbers: the weak gravitational lensing of the CMB. We showed that CMB lensing not only produces signal confusion for non-Gaussianity searches, it also produces large extra variance. Thus far, the effect of non-Gaussian covariance from lensing has not had a major impact on constraints, impacting Planck at most at the $10\%$ level. We showed that for future experiments, whose sensitivities and resolutions enable measurements up to $\ell_{\rm max} \sim 5000$, will be affected significantly. By far the largest impact is on local non-Gaussianities, where lensing will introduce an increase in the standard deviation of up to $35\%$ for an experiment like SO, largely removing all gains obtained over Planck. 

Fortunately, the effect can be mitigated. We explored two different avenues. First, we investigated the possibility of removing this effect by ISW-lensing marginalization. While in theory this should work well, the mitigation was less effective when tested in simulations. We believe this arises from the higher order terms which were neglected in the theoretical calculation. Second, we explored the possibility of delensing. Delensing was found to work very effectively, and for an SO-like experiment this resulted in an almost complete removal of the extra lensing variance. For the case of no primordial non-Gaussainity, delensing does not introduce any biases (i.e., it does not introduce a bispectrum). We find that the bias to a non-zero signal will be at the $10\%$ level, but that this can be avoided by reconstructing the lensing potential with combinations that include a $B$-mode map. For experiments beyond SO this can be done with minimal cost to the lensing reconstruction. This presents a new and compelling case to perform delensing of $T$ and $E$ maps. While the aim of this paper was to discuss the extra variance induced by lensing, delensing these maps before a measurement will also remove bispectra sourced by secondaries, such as those induced by ISW/reionizaton - lensing correlations. Results presented here are directly relevent for SO and the proposed CMB-S4 experiment. 

\acknowledgements

The authors are very grateful for useful discussions with Anthony Challinor, Antony Lewis, and Toshiya Namikawa. The NERSC and TIGRESS computing facilities were used in this work. P.D.M. acknowledges support from Senior Kavli
Institute Fellowships at the University of Cambridge (where this work was initiated) and the Netherlands organization for scientific research (NWO) VIDI grant (dossier 639.042.730). W.R.C. acknowledges support from the UK Science and Technology Facilities Council (grant number ST/N000927/1). A.v.E. would like to thank the Kavli Institute for Cosmology Cambridge for their support and hospitality for his visit under their Visiting Scholars program, during which some of this work was performed. This work made use of the TIGER cluster at Princeton and used resources of the National Energy Research Scientific Computing Center (NERSC), a U.S. Department of Energy Office of Science User Facility operated under Contract No. DE-AC02-05CH11231.

\begin{widetext}

\appendix

\section{Lensing induced covariance} \label{app:lensingVariance}
Here we show some of the details of the calculations that went into the derivation of the lensing induced covariance. 
\subsection{Flat sky} \label{app:flatSkyderivation}

Firstly we can write the Fourier coefficients as 
\be
T(\bl_i) = \tilde{T}_i + Q_i
\ee
where to linear order in the lensing potential
\be
Q_i =\dtl{1}\tilde{T}(\bl'_1)\phi(\bl_i-\bl'_1)(\bl_i-\bl'_1)\cdot\bl'_1.
\ee
Now, let us define the total covariance:

\be
{\rm Var}(\widehat{B}) \equiv \langle \widehat{B} \widehat{B}^* \rangle = 4\pi^2\delta^{(2)}(\bl_1+\bl_2+\bl_3) 4\pi^2\delta^{(2)}(\bl_1'+\bl_2'+\bl_3') \langle T(\bl_1)T(\bl_2)T(\bl_3)T^*(\bl_1')T^*(\bl_2')T^*(\bl_3')\rangle.
\label{eq:totalCov}
\ee
The additional covariance from lensing requires us to compute the connected lensing 4-point function.
We work to lowest order in $\phi$ and assume that lensing potential is not correlated with the unlensed temperature field (i.e., no ISW-lensing). We then have that the connected trispectrum is given by
\be
\langle T(\bl_2) T(\bl_3) T^*(\bl_2') T^*(\bl_3')\rangle_{\rm c} &=& 
 \langle \langle \tilde{T}_2 Q_{2'}^*\rangle_T \langle \tilde{T}_3 Q_{3'}^*\rangle_T \rangle_{\phi} + 11\; {\rm perm}.\nonumber \\
\ee

We introduced double bracket notation to show that either the (unlensed) temperature fields need to be contracted or the potentials.  Let us compute one such term for clarity and then use symmetry to obtain all other terms

\be
\langle \langle \tilde{T}_2 Q_{2'}^*\rangle_T \langle \tilde{T}_3 Q_{3'}^*\rangle_T \rangle_{\phi} &\equiv& \int \frac{\mathrm{d}^2\bl}{(2\pi)^2} \frac{\mathrm{d}^2\bl'}{(2\pi)^2} \langle \tilde{T}(\bl_2) \tilde{T}^*(\bl)\rangle \langle \tilde{T}(\bl_3) \tilde{T}^*(\bl')\rangle \langle \phi^*(\bl_{2}'-\bl)(\bl_{2}'-\bl)\cdot\bl \phi^*(\bl_{3}'-\bl')(\bl_{3}'-\bl')\cdot\bl'\rangle \nonumber \\
&=& (2\pi)^2\delta(\bl_2 + \bl_3 - \bl_{2}' - \bl_{3}') \tilde{C}_{\ell_2} \tilde{C}_{\ell_3} C_{|\bl_2 - \bl_{2'}|}^{\phi\phi} (\bl_2 - \bl_{2}') \cdot \bl_2 (\bl_3 - \bl_{3}') \cdot \bl_3
\label{eq:FS4pt}
\ee
There are 11 permutations of this term and those are indeed identical to the terms found by \cite{Zaldarriaga2000}.  In the bispectrum covariance there are then 6!/4!/2, i.e., 15, possibilities for the choice of which two fields to contract in the power spectrum term. However, six of those correspond to a zero lag or mean value, which is explicitly set to zero. Only nine permutations persist, multiplied by 12 permutations from the connected 4-point function in Eq.~\eqref{eq:FS4pt}.

We identify three distinguishable terms when computing the covariance, i.e., 
\be
\langle \langle \tilde{T}_2 Q_3\rangle_T \langle \tilde{T}^*_{2'} Q_{3'}^*\rangle_T \rangle_{\phi} & =(2\pi)^2 \delta(\bl_2 + \bl_3 - \bl_2' - \bl_3')\tilde{C}_{\ell_2} \tilde{C}_{\ell_2'} C^{\phi\phi}_{|\bl_2+\bl_3|}(\bl_3+\bl_2)\cdot\bl_2(\bl_3'+\bl_2')\cdot\bl_2' \\
\langle \langle \tilde{T}_2 Q_{3'}^*\rangle_T \langle \tilde{T}_{2'}^* Q_{3}\rangle_T \rangle_{\phi} & = (2\pi)^2\delta(\bl_2 + \bl_3 - \bl_2' - \bl_3')\tilde{C}_{\ell_2} \tilde{C}_{\ell_2'} C^{\phi\phi}_{|\bl_2-\bl_3'|}(\bl_2 -\bl_3')\cdot\bl_2(\bl_2'-\bl_3)\cdot\bl_2' \\
\langle \langle \tilde{T}_2 Q_{2'}^*\rangle_T \langle \tilde{T}_{3} Q_{3'}^*\rangle_T \rangle_{\phi} & = (2\pi)^2\delta(\bl_2 + \bl_3 - \bl_2' - \bl_3')\tilde{C}_{\ell_2} \tilde{C}_{\ell_3} C^{\phi\phi}_{|\bl_2-\bl_{2}'|}(\bl_2-\bl_{2}')\cdot\bl_2(\bl_3-\bl_{3}')\cdot\bl_3
\ee

Note that these are all just permutations on the term in Eq. \eqref{eq:FS4pt}, however they contribute differently to the bispectrum due to triangle conditions enforced on the bispectrum. In particular, we find that the first contribution is significantly larger than the other two terms. Combining Eq.~\eqref{eq:FS4pt} with Eq.~\eqref{eq:totalCov}, collecting terms, applying delta functions and counting permutations we find 
\be
{\rm Var}(\widehat{B})_{2p\times4p} &=& (2\pi)^8\delta^{(2)}(0)\delta^{(2)}(\bl_1-\bl_1')\delta^{(2)}(\sum\limits_i \bl_i)\delta^{(2)}(\sum\limits_i \bl_i') C_{\ell_1} C_{\ell_1}^{\phi\phi} \bl_1\cdot \bl_2 \tilde{C}_{\ell_2} \bl_1\cdot \bl_2' \tilde{C}_{\ell_2'} +{\rm 35\;perm.} \nonumber \\
& &+ (2\pi)^6\delta^{(2)}(\bl_1-\bl_1') \delta^{(2)}(\sum\limits_i \bl_i)\delta^{(2)}(\sum\limits_i \bl_i')C_{\ell_1} \int \mathrm{d}^2\bm{\hat{n}} C_{|\bl_2-\bl_3'|}^{\phi\phi} G(\bl_2,\bl_3')\tilde{C}_{\ell_2} G^*(\bl_2',\bl_3)\tilde{C}_{\ell_2'} +{\rm 35\;perm.} \nonumber \\
& &+ (2\pi)^6\delta^{(2)}(\bl_1-\bl_1')\delta^{(2)}(\sum\limits_i \bl_i)\delta^{(2)}(\sum\limits_i \bl_i')C_{\ell_1} \int \mathrm{d}^2\bm{\hat{n}} C_{|\bl_2-\bl_2'|}^{\phi\phi} G(\bl_2,\bl_2')\tilde{C}_{\ell_2} G(\bl_3,\bl_3')\tilde{C}_{\ell_3} +{\rm 35\;perm.}
\ee
Here we have defined 
\be
G(\bl_i,\bl_j) &=& \bl_i \cdot ( \bl_i - \bl_j) e^{i \bm{\hat{n}} \cdot (\bl_i-\bl_j)},
\ee
and we used 
\be
\delta(\bl_i+\bl_j) = \int \frac{\mathrm{d}^2\bm{\hat{n}}}{(2\pi)^2} e^{i (\bl_i+\bl_j) \cdot \bm{\hat{n}}}.
\label{eq:deltaFuncRexpress}
\ee

Note that we used Eq. \eqref{eq:deltaFuncRexpress} to express the second and third terms with an integral over the line of sight direction. This highlights the similarity with the full-sky result (which involves an extra summation). 
From the perspective of the estimator variance to be computed in Sec.~\ref{sec:sizeOfEffect}, when summing over all triplets the 36 permutations just introduce a multiplicity.

\subsection{Full sky}\label{app:fullSkyderivation}
We start with Eq.~\eqref{eq:BisCovFullSky} and compute the covariance term by term. For convenience we repeat that equation here
\be
\Delta{\rm Var}(\widehat{B})_{2p\times4p} &=& \sum\limits_{mm'} \left( \begin{matrix} \ell_1 & \ell_2 & \ell_3 \\ m_1 & m_2 & m_3 \end{matrix} \right) \left( \begin{matrix} \ell_1' & \ell_2' & \ell_3' \\ m_1' & m_2' & m_3'\end{matrix} \right) \delta_{\ell_1 \ell_1'} \delta_{m_1 m_1'} C_{\ell_1} \left[\overbrace{\langle \tilde{a}_{\ell_2 m_2}\tilde{a}_{\ell_3 m_3} Q_{\ell_2'm_2'}^*Q^*_{\ell_3'm_3'}\rangle_{\rm c}}^{(1)} + \right. \nonumber \\
&& \left. \overbrace{\langle \tilde{a}_{\ell_2m_2}Q_{\ell_3m_3}\tilde{a}^{*}_{\ell_2'm_2'}Q^*_{\ell_3' m_3'}\rangle_{\rm c}}^{(2)} \right] + {\rm perm} +\mathcal{O}(\phi^3)
\ee
Let us first compute terms in (1):
\be
(1) &=& \tilde{C}_{\ell_2} \tilde{C}_{\ell_3} \sum\limits_{LM} C_{L}^{\phi\phi} (-1)^{m_2}(-1)^{m_3} (-1)^M\left[\left( \begin{matrix} \ell_2' & L & \ell_2 \\ m_2' & M & -m_2 \end{matrix} \right) \left( \begin{matrix} \ell_3' & L & \ell_3 \\ m_3' & -M & -m_3 \end{matrix} \right) F_{\ell_2' L \ell_2}F_{\ell_3' L \ell_3} + \right.\nonumber \\
&& \left.\left( \begin{matrix} \ell_2' & L & \ell_3 \\ m_2' & M & -m_3 \end{matrix} \right) \left( \begin{matrix} \ell_3' & L & \ell_2 \\ m_3' & -M & -m_2 \end{matrix} \right) F_{\ell_2' L \ell_3}F_{\ell_3' L \ell_2} \right]. 
\ee
Here we used $a_{\ell m} = (-1)^m a^*_{\ell -m}$. Note that contracting $a_{\ell_2 m_2}$ with $a_{\ell_3m_3}$, via
\be
(-1)^m \left( \begin{matrix} \ell & \ell & \ell' \\ m & -m & 0 \end{matrix} \right) = \frac{(-1)^{\ell}}{\sqrt{2\ell+1}}\delta_{\ell'0}
\ee
would force $\ell_3$ to be 0, and hence it is not allowed. 

For the second term we have 
\be
(2) &=&\tilde{C}_{\ell_2} \tilde{C}_{\ell_2'} \sum\limits_{LM} C_{L}^{\phi\phi} \left[ (-1)^{m_2}(-1)^{m_2'}\left( \begin{matrix} \ell_3 & L & \ell_2' \\ m_3 & M & -m_2' \end{matrix} \right) \left( \begin{matrix} \ell_3' & L & \ell_2 \\ m_3' & M & -m_2 \end{matrix} \right) F_{\ell_3' L \ell_2}F_{\ell_3 L \ell_2'} + \right.\nonumber \\
&& \left.\left( \begin{matrix} \ell_3 & L & \ell_2 \\ m_3 & M & m_2 \end{matrix} \right) \left( \begin{matrix} \ell_3' & L & \ell_2' \\ m_3' & M & m_2' \end{matrix} \right) F_{\ell_3 L \ell_2}F_{\ell_3' L \ell_2'} \right] .\ee
Not counting the Wigner 3-$j$ symbols in the coupling matrix $F$, we generally encounter 4 Wigner 3-$j$'s, summed over 6 indices (5 $m$'s and one $L$). To simplify these remaining terms we make generously use of the following identity 
\be
\sum\limits_{m_1 m_2 m_4 m_5 m_6} (-1)^{\sum\limits_i^6 \ell_i-m_i } (-1)^{-\ell_3+m_3} && \left( \begin{matrix} \ell_2 & \ell_3 & \ell_1 \\ m_2 & -m_3 & m_1 \end{matrix} \right)\left( \begin{matrix} \ell_1 & \ell_5 & \ell_6 \\ -m_1 & m_5 & m_6 \end{matrix} \right)
\left( \begin{matrix} \ell_5 & \ell_3' & \ell_4 \\ -m_5 & m_3' & m_4 \end{matrix} \right)
\left( \begin{matrix} \ell_4 & \ell_2 & \ell_6 \\ -m_4 & -m_2 & -m_6 \end{matrix} \right)\nonumber\\
&&=\frac{(-1)^{\ell_3-m_3}}{2\ell_3+1}\delta_{\ell_3\ell_3'} \delta_{m_3m_3'}\left\{ \begin{matrix} \ell_1 & \ell_2 & \ell_3 \\ \ell_4 & \ell_5 & \ell_6 \end{matrix} \right\}
\ee
The identity contains a Wigner 6-$j$ symbol, but allows us to rewrite all terms above in just sums over $L$ (removing all visible sums over $m$ and 4 Wigner 3-$j$ symbols). 

Let us start with the first line of (1). We have the following sum 
\be
\sum\limits_{m_1 m_2 m_3 m_2'm_3'M} (-1)^{m_2+m_3+M} \left( \begin{matrix} \ell_1 & \ell_2 & \ell_3 \\ m_1 & m_2 & m_3 \end{matrix} \right)\left( \begin{matrix} \ell_1 & \ell_2' & \ell_3' \\ m_1 & m_2' & m_3' \end{matrix} \right)\left( \begin{matrix} \ell_2' & L & \ell_2 \\ m_2' & M & -m_2 \end{matrix} \right) \left( \begin{matrix} \ell_3' & L & \ell_3 \\ m_3' & -M & -m_3 \end{matrix} \right)
\ee
We can massage this into the identity above, by identifying $\ell_2'\rightarrow \ell_5$, $\ell_3'\rightarrow \ell_6$ and $L \rightarrow \ell_4$ (and similar for the $m$'s) We then also reverse sign of $m_5$ and $m_6$ (which is allowed because these are just dummies). We then have 
\be
\sum\limits_{m_1 m_2 m_3 m_4 m_5 m_6} (-1)^{m_2+m_3+m_4} \left( \begin{matrix} \ell_1 & \ell_2 & \ell_3 \\ m_1 & m_2 & m_3 \end{matrix} \right)\left( \begin{matrix} \ell_1 & \ell_5 & \ell_6 \\ m_1 & -m_5 & -m_6 \end{matrix} \right)\left( \begin{matrix} \ell_5 & \ell_4 & \ell_2 \\ -m_5 & m_4 & -m_2 \end{matrix} \right) \left( \begin{matrix} \ell_6 & \ell_4 & \ell_3 \\ -m_6 & -m_4 & -m_3 \end{matrix} \right)\nonumber
\ee
We can apply permutations of the columns picking up $(-1)^{\sum \ell}$ for odd permutations, and sign inversion of the $m$'s picks up a similar factor. After some manipulation we have
\be
\sum\limits_{m_1 m_2 m_3 m_4 m_5 m_6} (-1)^{m_2+m_3+m_4}(-1)^{\ell_1+\ell_2+\ell_4+\ell_6} \left( \begin{matrix} \ell_3 & \ell_2 & \ell_1 \\ m_3 & m_2 & m_1 \end{matrix} \right)\left( \begin{matrix} \ell_1 & \ell_5 & \ell_6 \\ -m_1 & m_5 & m_6 \end{matrix} \right)\left( \begin{matrix} \ell_5 & \ell_2 & \ell_4 \\ -m_5 & -m_2 & m_4 \end{matrix} \right) \left( \begin{matrix} \ell_4 & \ell_3 & \ell_6 \\ -m_4 & -m_3 & -m_6 \end{matrix} \right)\nonumber
\ee
We then switch dummy labels $3$ and $2$ and invert the sign of $m_3$
\be
\sum\limits_{m_1 m_2 m_3 m_4 m_5 m_6} (-1)^{m_2+m_3+m_4}(-1)^{\ell_1+\ell_3+\ell_4+\ell_6} \left( \begin{matrix} \ell_2 & \ell_3 & \ell_1 \\ m_2 & -m_3 & m_1 \end{matrix} \right)\left( \begin{matrix} \ell_1 & \ell_5 & \ell_6 \\ -m_1 & m_5 & m_6 \end{matrix} \right)\left( \begin{matrix} \ell_5 & \ell_3 & \ell_4 \\ -m_5 & m_3 & m_4 \end{matrix} \right) \left( \begin{matrix} \ell_4 & \ell_2 & \ell_6 \\ -m_4 & -m_2 & -m_6 \end{matrix} \right)\nonumber
\ee
The final step is to realise that $(-1)^m = (-1)^{-m}$ and that the second Wigner 3-$j$ sets $m_1 = m_5+m_6 \rightarrow m_1+m_5+m_6 = 2m_1$ which is an even number. We can thus write 
\be
\sum\limits_{m_3} (-1)^{-\ell_2+\ell_3+\ell_5+m_3} \sum\limits_{m_1 m_2 m_4 m_5 m_6} (-1)^{\sum\limits_i^6 \ell_i-m_i } (-1)^{-\ell_3+m_3} \left( \begin{matrix} \ell_2 & \ell_3 & \ell_1 \\ m_2 & -m_3 & m_1 \end{matrix} \right)\left( \begin{matrix} \ell_1 & \ell_5 & \ell_6 \\ -m_1 & m_5 & m_6 \end{matrix} \right) \times \nonumber \\ \left( \begin{matrix} \ell_5 & \ell_3 & \ell_4 \\ -m_5 & m_3 & m_4 \end{matrix} \right) \left( \begin{matrix} \ell_4 & \ell_2 & \ell_6 \\ -m_4 & -m_2 & -m_6 \end{matrix} \right)\nonumber.
\ee
For the inner sum we can use this identity and the first correction to the variance becomes (after substituting back the correct $\ell$) 
\be
\Delta{\rm Var}(\widehat{B})_{2p\times4p}[1] &=& \delta_{\ell_1\ell_1'}C_{\ell_1} \tilde{C}_{\ell_2} \tilde{C}_{\ell_3} (-1)^{\ell_3+\ell_2'}\sum\limits_L C_L^{\phi\phi}\left\{ \begin{matrix} \ell_1 & \ell_3 & \ell_2 \\ L & \ell_2' & \ell_3' \end{matrix} \right\}F_{\ell_2'L\ell_2}F_{\ell_3'L\ell_3} + {\rm 3\; perm.}
\ee

Next, we compute the second line of (1). We have 
\be
\sum\limits_{m_1 m_2 m_3 m_2'm_3'M} (-1)^{m_2+m_3+M} \left( \begin{matrix} \ell_1 & \ell_2 & \ell_3 \\ m_1 & m_2 & m_3 \end{matrix} \right)\left( \begin{matrix} \ell_1 & \ell_2' & \ell'_3 \\ m_1 & m_2' & m'_3 \end{matrix} \right)\left( \begin{matrix} \ell_2' & L & \ell_2 \\ m_2' & M & -m_3 \end{matrix} \right) \left( \begin{matrix} \ell_3' & L & \ell_2 \\ m_3' & -M & -m_2 \end{matrix} \right),
\ee
and make the same identifications as before , and changing sign of $m_3$, $m_4$, $m_5$ and $m_6$ we find 
\be
\sum\limits_{m_3} (-1)^{\ell_3-\ell_2 -\ell_5-m_3} \sum\limits_{m_1 m_2 m_4 m_5 m_6} (-1)^{\sum\limits_i^6 \ell_i-m_i } (-1)^{-\ell_3+m_3} \left( \begin{matrix} \ell_2 & \ell_3 & \ell_1 \\ m_2 & -m_3 & m_1 \end{matrix} \right)\left( \begin{matrix} \ell_1 & \ell_5 & \ell_6 \\ -m_1 & m_5 & m_6 \end{matrix} \right) \times \nonumber \\ \left( \begin{matrix} \ell_5 & \ell_3 & \ell_4 \\ -m_5 & m_3 & m_4 \end{matrix} \right) \left( \begin{matrix} \ell_4 & \ell_2 & \ell_6 \\ -m_4 & -m_2 & -m_6 \end{matrix} \right)\nonumber
\ee
The second correction correction to the covariance from term one thus becomes
\be
\Delta{\rm Var}(\widehat{B})_{2p\times4p}[2] &=& \delta_{\ell_1\ell_1'}C_{\ell_1} \tilde{C}_{\ell_2} \tilde{C}_{\ell_3} (-1)^{-\ell_2-\ell'_2}\sum\limits_L C_L^{\phi\phi}\left\{ \begin{matrix} \ell_1 & \ell_2 & \ell_3 \\ L & \ell_2' & \ell_3' \end{matrix} \right\}F_{\ell_2'L\ell_3}F_{\ell_3'L\ell_2} + {\rm 1\;perm.}
\ee

Let us continue with the first line of the second term, i.e., 
\be
\sum\limits_{m_1 m_2 m_3 m_2'm_3'M} (-1)^{m_2+m_2'} \left( \begin{matrix} \ell_1 & \ell_2 & \ell_3 \\ m_1 & m_2 & m_3 \end{matrix} \right)\left( \begin{matrix} \ell_1 & \ell_2' & \ell_3 \\ m_1 & m_2' & m_3 \end{matrix} \right)\left( \begin{matrix} \ell_3 & L & \ell_2' \\ m_3 & M & -m_2' \end{matrix} \right) \left( \begin{matrix} \ell_3' & L & \ell_2 \\ m_3' & M & -m_2 \end{matrix} \right).
\ee
We again make the same identification in $\ell$ as well as changing sign of $m_3$, $m_4$, $m_5$ and $m_6$ we find 
\be
\sum\limits_{m_3} (-1)^{\ell_3-\ell_2 -\ell_5+m_3} \sum\limits_{m_1 m_2 m_4 m_5 m_6} (-1)^{\sum\limits_i^6 \ell_i-m_i } (-1)^{-\ell_3+m_3} \left( \begin{matrix} \ell_2 & \ell_3 & \ell_1 \\ m_2 & -m_3 & m_1 \end{matrix} \right)\left( \begin{matrix} \ell_1 & \ell_5 & \ell_6 \\ -m_1 & m_5 & m_6 \end{matrix} \right) \times \nonumber \\ \left( \begin{matrix} \ell_5 & \ell_3 & \ell_4 \\ -m_5 & m_3 & m_4 \end{matrix} \right) \left( \begin{matrix} \ell_4 & \ell_2 & \ell_6 \\ -m_4 & -m_2 & -m_6 \end{matrix} \right)\nonumber
\ee
and hence 
\be
\Delta{\rm Var}(\widehat{B})_{2p\times4p}[3] &=& \delta_{\ell_1\ell_1'}C_{\ell_1} \tilde{C}_{\ell_2} \tilde{C}_{\ell_2'} (-1)^{-\ell_2-\ell_2'}\sum\limits_L C_L^{\phi\phi}\left\{ \begin{matrix} \ell_1 & \ell_2 & \ell_3 \\ L & \ell_2' & \ell_3' \end{matrix} \right\}F_{\ell_3'L\ell_2}F_{\ell_3L\ell_2'} + {\rm 1\; perm.}
\ee
The second line in (2) reads 
\be 
\sum\limits_{m_1 m_2 m_3 m_2'm_3'M}\left( \begin{matrix} \ell_1 & \ell_2 & \ell_3 \\ m_1 & m_2 & m_3 \end{matrix} \right)\left( \begin{matrix} \ell_1 & \ell_2' & \ell_3 \\ m_1 & m_2' & m_3 \end{matrix} \right)\left( \begin{matrix} \ell_3 & L & \ell_2 \\ m_3 & M & m_2 \end{matrix} \right) \left( \begin{matrix} \ell_3' & L & \ell_2' \\ m_3' & M & m_2' \end{matrix} \right).
\ee
Reordering of columns allows us to use a simple identity, this term becomes $\sum\limits_{m} \frac{\delta_{\ell_1 L} \delta_{m_1 M}}{2\ell_1+1}$, 
\be \label{eq:DB4}
\Delta{\rm Var}(\widehat{B})_{2p\times4p}[4] & = & \delta_{\ell_1\ell_1'}C_{\ell_1} \tilde{C}_{\ell_2} \tilde{C}_{\ell_2'} C_{\ell_1}^{\phi\phi}F_{\ell_3 \ell_1 \ell_2}F_{\ell_3' \ell_1 \ell_2'}
\frac{1}{2\ell_1+1}+{\rm 3\;perm.}
\ee
This completes all the terms that you can expect in the covariance.

\section{Examining the non-leading contributions} \label{app:N1check}
In our theoretical modelling we identified one of the terms as the leading term, Eq. \eqref{eq:DB4}, and assumed that the other terms are subdominant. This assumption was motivated by the fact that these terms have a similar form to those seen in CMB lensing reconstruction analyses, where they have been found to be small \citep{Kesden2003}. To validate this assumption we used the methods developed in \cite{Story2015} to isolate this term in simulations. Here we summarise this method and refer the reader to \cite{Story2015} for more details. As was seen in Appendix \ref{app:lensingVariance}, the non-leading terms all arise from mixed between the two bispectra, i.e., couplings between the primed and unprimed indices. To isolate the sub-leading variance term we compute the difference in variance of two bispectra measurements. The first bispectrum is measured on three different realisations of the primary CMB that have been lensed by the same lensing potential, hereafter these maps are denoted $D^{\phi}_{\ell,m}$, $E^{\phi}_{\ell,m}$ and $F^{\phi}_{\ell,m}$. The second bispectrum is measured on three different realisations of the primary CMB that have been lensing by different lensing potentials, hereafter we denote these maps by $D^{\phi_D}_{\ell,m}$, $E^{\phi_E}_{\ell,m}$ and $F^{\phi_F}_{\ell,m}$. We then examine the difference of their variances, i.e.
\be
 \mathrm{Var}(\widehat{B}_1)-\mathrm{Var}(\widehat{B}_2) =\langle \left( \sum \limits_{\ell_i,m_i} \left( \begin{matrix} \ell_1 & \ell_2 & \ell_3 \\ m_1 & m_2 & m_3 \end{matrix} \right) D^{\phi}_{\ell_1,m_1}E^{\phi}_{\ell_2,m_2}F^{\phi}_{\ell_3,m_3}\right)^2- \left( \sum \limits_{\ell_i,m_i} \left( \begin{matrix} \ell_1 & \ell_2 & \ell_3 \\ m_1 & m_2 & m_3 \end{matrix} \right) D^{\phi_D}_{\ell_1,m_1}E^{\phi_E}_{\ell_2,m_2}F^{\phi_F}_{\ell_3,m_3} \right)^2\rangle. \nonumber \\
\ee
Computing this variance to linear order in $C^{\phi\phi}$ we find that it contains all of the terms in the full expression, Eq. ~\eqref{eq:BisCovFullSkyEval}, except the leading term, Eq.~\eqref{eq:DB4}, as we wanted. As computing this for both temperature and polarisation would be computationally expensive, we computed this only for the temperature case. The polarization case is analogous and so we expect that the non-leading terms should contribute to the total variance at a similar level. Using this configuration we measured the contribution of the non-leading terms using 342 of the lensing CMB simulations for a cosmic variance limited setup. We find, after accounting for multiplicity, these terms contribute $13.5\%$ of the total variance and hence are subdominant.

\section{Analytic calculation of the delensing bias} \label{app:delensingBias}

As discussed in Section \ref{sec:delensing}, delensing introduces a multiplicative bias to primordial non-Gaussianity measurements. Here we describe the calculation of this bias to leading order in $C^{\phi \phi}$.
After delensing the spherical harmonic coefficients can be written as
\begin{align}
a^{T,\rm delensed}_{\ell,m} &= {a}^{T}_{\ell,m}- \sum\limits_{L,M}\sum \limits_{\ell',m'} W_L\hat{\phi}^*_{L,M}{a^{T^*}}_{\ell',m'} \left( \begin{matrix} \ell & L & \ell' \\ m & M & m' \end{matrix} \right) F_{\ell,L,\ell'}, \nonumber \\
a^{E,\rm delensed}_{\ell,m} & ={a}^{E}_{\ell,m} - \sum \limits_{L,M}\sum \limits_{\ell',m'} \left( \begin{matrix} \ell & L & \ell' \\ m & M & m' \end{matrix} \right) W_L\hat{\phi}^*_{L,M}\left[ F^{+2}_{\ell,L,\ell'} {{a}^{E^*}}_{\ell',m'}-i F^{-2}_{\ell,L,\ell'}{{a}^{B^*}}_{\ell',m'} \right], \nonumber \\
a^{B,\rm delensed}_{\ell,m} &= {a}^{B}_{\ell,m} -\sum \limits_{L,M}\sum \limits_{\ell',m'} \left( \begin{matrix} \ell & L & \ell' \\ m & M & m' \end{matrix} \right) W_L\hat{\phi}^*_{L,M}\left[ F^{+2}_{\ell,L,\ell'} {{a}^{B^*}}_{\ell',m'}+i F^{-2}_{\ell,L,\ell'}{{a}^{E^*}}_{\ell',m'}\right], 
\end{align}
where $\hat{\phi}$ is the reconstructed lensing potential and $W_L = {C^{\phi\phi}_{L}}/{[C^{\phi\phi}_{L}+N^{\phi\phi}_{L}]}.$ We use the quadratic estimator to reconstruct the lensing potential, following \cite{Okamoto2003}, and thus have
\begin{align}
\hat{\phi}_{L,M} = A_L \sum\limits_{\ell_a,\ell_b} \left( \begin{matrix} \ell_a & \ell_b &L \\ m_a & m_b & M \end{matrix} \right) g^{\alpha,\beta}_{\ell_a,\ell_b,L}a^{\alpha^*}_{\ell_a,m_a}a^{\beta^*}_{\ell_b,m_b},
\end{align}
where $A_L$ is the normalization and $g^{\alpha,\beta}_{\ell_a,\ell_b,L}$ are the lensing reconstruction weights and the indices $\alpha$ and $\beta$ are summed over maps used in the reconstruction. Using these we can compute the leading order effect of delensing on the bispectrum. We only consider scalar bispectra and so we do not consider bispectra involving $B$-mode fields. 
\begin{align} \label{eq:delensedBI}
\hat{f}_{NL}^{\rm delensed} &=\frac{1}{N} \sum \tilde{B}^{X_1,X_2,X_3}_{\ell_1,\ell_2,\ell_3} \left( \begin{matrix} \ell_1 & \ell_2 & \ell_3 \\ m_1 & m_2 & m_3 \end{matrix} \right) \langle a^{X_1,\rm delensed}_{\ell_1,m_1} a^{X_2,\rm delensed}_{\ell_2,m_2} a^{X_3,\rm delensed}_{\ell_3,m_3} \rangle \nonumber \\ &
=\frac{1}{N}\sum \limits_{S_3}\limits \tilde{B}^{X_1,X_2,X_3}_{\ell_1,\ell_2,\ell_3} \left( \begin{matrix} \ell_1 & \ell_2 & \ell_3 \\ m_1 & m_2 & m_3 \end{matrix} \right) \langle \tilde{a}^{X_1}_{\ell_1,m_1}\tilde{a}^{X_2}_{\ell_2,m_2}\left(\tilde{a}^{X_3}_{\ell_3,m_3}-3\sum\limits_{L,\ell'} W_L\hat{\phi}^*_{L,M}{{a}^{S_3*}}_{\ell',m'} \left( \begin{matrix} \ell_3& L & \ell' \\ m _3& M & m' \end{matrix} \right) F^{S_3}_{\ell_3,L,\ell'}\right) \rangle, 
\end{align}
where $ S_i$ is T if $X_i$ = T and $S_i \in [E,B]$ otherwise, 
\be
F^{S_i}_{\ell_i,L,\ell'}= 
\begin{cases}
F^{0}_{\ell_i,L,\ell'} \text{ if $S_i =T$} \\
F^{2}_{\ell_i,L,\ell'} \text{ if $S_i =E$} \\
-iF^{-2}_{\ell_i,L,\ell'} \text{ if $S_i =B$},
\end{cases}
\ee
and, for conciseness, we defined
\be
 \tilde{B}^{X_1,X_2,X_3}_{\ell_1,\ell_2,\ell_3} = \sum {C^{-1}}^{X_1,X_1'}_{\ell_1} {C^{-1}}^{X_2,X_2'}_{\ell_2} {C^{-1}}^{X_3,X_3'}_{\ell_3} {B}^{X_1',X_2',X_3'}_{\ell_1,\ell_2,\ell_3}.
\ee
Now we evaluate Eq.~\eqref{eq:delensedBI} for the case of non zero $f_{\rm NL}$, i.e., when
\be
\langle a^{X_1}_{\ell_1,m_1}a^{X_2}_{\ell_2,m_2}a^{X_3}_{\ell_3,m_3} \rangle = f_{\rm NL} B^{X_1,X_2,X_3}_{\ell_1,\ell_2,\ell_3} \left( \begin{matrix} \ell_1 & \ell_2 & \ell_3 \\ m_1 & m_2 & m_3 \end{matrix} \right).
\ee
We find
\begin{align}
\langle \hat{f}_{\rm NL} \rangle = f_{\rm NL}\bigg( & 1-\frac{6}{N} \sum \tilde{B}^{X_1,X_2,X_3}_{\ell_1,\ell_2,\ell_3}C^{X_2,S_3}_{\ell_2} F^{S_3}_{\ell_3,\ell_1,\ell_2}W_{\ell_1}A_{\ell_1} B^{X_1,\alpha,\beta}_{\ell_1,\ell_a,\ell_b}{g^{\alpha,\beta}}^*_{\ell_a,\ell_b,\ell_1} \frac{1}{2\ell_1+1} \nonumber \\
&-\frac{6}{N} \sum \tilde{B}^{X_1,X_2,X_3}_{\ell_1,\ell_2,\ell_3} B^{X_1,X_2,\beta}_{\ell_1,\ell_2,\ell_3}C^{S_3,\alpha}_{\ell'} F^{S_3}_{\ell_3,L,\ell'}W_L A_{L} {g^{\alpha\beta} }^*_{\ell',\ell_3,L}\frac{1}{2\ell_3+1}\nonumber \\
& -\frac{6}{N} \sum \tilde{B}^{X_1,X_2,X_3}_{\ell_1,\ell_2,\ell_3} B^{X_1,S_3,\beta}_{\ell_1,\ell',\ell_b}C^{X_2,\alpha}_{\ell'} F^{S_3}_{\ell_3,L,\ell'} W_LA_{L} {g^{\alpha\beta}}^*_{\ell_2,\ell_b,L}\frac{(-1)^{\ell_3+L}}{2\ell_1+1}\left\{ \begin{matrix} \ell_3 & \ell_2 & \ell_1 \\ \ell_b &\ell' & L \end{matrix} \right\} \bigg).
\label{eq:appBiasFull}
\end{align}
Thus, at leading order, we find that the bias is a multiplicative bias. We estimate the size of this bias by computing the first two terms. Unfortunately the last term is computationally intractable, but it is expected to be small as it has the same structure as the sub-leading variance terms (which we explicitly verified are small). This claim is further supported as in Section \ref{sec:delensing} we find that the first two terms in Eq. \eqref{eq:appBiasFull} predict a level of bias that is consistent with the bias measured in simulations.

\end{widetext}
\bibliographystyle{apsrev.bst}
\bibliography{project}
\end{document}